\documentclass{article}
\usepackage[utf8]{inputenc}
\usepackage[T1]{fontenc}
\usepackage[english]{babel}
\usepackage{amssymb}
\usepackage{amsmath}
\usepackage{amsfonts}
\usepackage{amssymb}
\usepackage{mathrsfs}
\usepackage{graphicx}
\usepackage{enumerate}
\usepackage{epstopdf}   
\usepackage{cite}
\usepackage{caption}
\usepackage{subcaption} 	

\usepackage{float}
\usepackage{hyperref}
\usepackage{geometry}
\title{Finite volume simulation of a semi-linear \\
Neumann problem (Keller-Segel model) \\
on rectangular domains}

\author{{Nardjess Benoudina$^{1}$}, Fatima Zohra Boutaf$^{2}$, Nasserdine Kechkar$^{3}$\\
{\small \it\  $^1$Department of
Mathematics, Zhejiang Normal University,} \\
{\small \it\ Jinhua 321004, PR China }  \\
{\small \it \ $^2$Department of Mathematics, University of Mohamed Boudiaf,}\\
{\small \it\  M'sila,  28000, Algeria}\\
{\small \it\  $^3$Department of Mathematics, Faculty of Exact Sciences,} \\
{\small \it\  University of Constantine $1$, }\\
{\small \it\  Constantine, 25017, Algeria} }

\begin{document}

\maketitle

\begin{abstract}
  In this study, the finite volume method is implemented for solving the problem of the semi-linear equation: %
$-d\, \Delta u+u=u^{q}$ ($d,q >0$) with a homogeneous Neumann boundary condition. This problem is equivalent to %
the known stationary Keller-Segel model, which arises in chemotaxis.After discretization, a nonlinear algebraic system %
is obtained and solved on the platform Matlab. As a result, many single-peaked and multi-peaked shapes in $3D$ and %
contour plots can be drawn depending on the parameters $d$ and $q$.
\end{abstract}

\section{Intoduction}
\indent In biology, chemotaxis is a type of cell movement that occurs when bodily cells, such as spermatozoa, the tube %
of pollen grains, bacteria, or other uni- or multicellular organisms, direct themselves or their movements in %
response to certain chemical species that are present in their environment. It is noteworthy to mention that %
chemotaxis plays a significant role in the development and physiological functioning of the organism %
\cite{Hillen2009,Meinhardt1982}. \\

In 1970, Keller and Segel proposed in \cite{Keller1970} a mathematical model in order to represent the process %
of amoebae transforming into chemotactic aggregates. They introduced a problem for a system of two semi-linear %
 PDEs for the amoeba population $w (x, t)$ and the chemical product concentration $v(x, t)$. This is given as follows:

\begin{equation} \label{kelSeg}
	\begin{cases}
	\;\;\quad \frac{\partial w}{\partial t}= D_{1}\Delta w-\chi \nabla .(w\nabla \phi (v)) \qquad x\in \Omega ,\; t>0,\\
           \;\;\quad\, \frac{\partial v}{\partial t}= D_{2}\Delta v+k(w,v) \qquad\qquad\quad\; x\in \Omega ,\; t>0,\\
	\;\,\quad \frac{\partial w}{\partial \overrightarrow{n}}=\frac{\partial v}{\partial\overrightarrow{n}}=0 \qquad\qquad\qquad\qquad\;\;\; x \in \partial\Omega ,\; t>0, \\
	w(x,0)=w_{0}(x)>0 \qquad\qquad\qquad\quad\;\;\, x \in \Omega,\\
	\; v(x,0)=v_{0}(x)>0 \qquad\qquad\qquad\qquad x \in \Omega,
          \end{cases} 
\end{equation}
where $\Omega $ is a bounded domain in $\mathbb{R}^{N}$ with a regular boundary $\partial \Omega$, $\phi$ %
a real function such as $\phi ^{^{\prime }}(r)>0$ for any $r>0$, $k(w,v)$ a real function with $k_{w}\geq 0$, %
$k_{v}\leq 0$ and $\overrightarrow{n}$ denotes the unit outer normal to $\partial \Omega $, whereas $D_{1}$, %
$D_{2}$ and $\chi $ are positive given constants. Here, as usual,
$$\Delta ={\sum_{i=1}^N }\frac{\partial ^{2}}{\partial x_{i}^{2}}, \quad \nabla =\left[ \frac{\partial }{\partial x_{1}} \; \frac{\partial }{\partial x_{2}} \; \hdots \; \frac{\partial }{\partial x_{N}} \right]^{T}.$$
A simple functional transformation reduces problem \eqref{kelSeg} in its stationary version to the equivalent %
problem for a single semi-linear PDE (see \cite{Lin1988}):
\begin{equation}\label{SemiLineaModl}
\left\{ 
\begin{array}{c}
-d \, \Delta u+u=u^{q}\ \ \ \  \text{in }\Omega \\ 
\frac{\partial u}{\partial \overrightarrow{n}}=0\ \ \ \ \ \ \ \ \ \ \ \ \ \
\ \ sur\text{ }\partial \Omega \\ 
u>0\text{ \ \ \ \ \ \ \ \ \ \ \ \ \ \ \ \ \ } \text{ for}\Omega%
\end{array}%
\right.
\end{equation}
with $d$ and $q$ being given positive constants. This problem arises in the investigation of steady-state solutions to %
certain reaction-diffusion systems involved in chemotaxis and morphogenesis. Therefore, it is widely studied to ensure %
the best understanding of this phenomenon. The existence and uniqueness of the least-energy solution to the problem %
\eqref{SemiLineaModl} have been proven in the literature (see, e.g., \cite{Grossi1999, Kwong1989, Lin1988}). However, many %
studies are focusing on the shape of the solution. In this respect, single-peaked and multi-peaked solutions are theoretically found in %
\cite{Ackermann1998, CAO1999, Ni1991, Wang1992} and predicting their locations \cite{Ni1993}. In addition, the %
boundary spike layer solutions are obtained and studied in \cite{Lee2020, Ni1998, Wei1998}. A numerical study based on %
the fast Fourier solver has been applied in \cite{Hwang2002} to the problem \eqref{SemiLineaModl} in order to investigate various solution %
forms. \\

The finite volume method is a well-adapted discretization technique for various types of simulation of conservation laws %
in elliptic, parabolic, hyperbolic, and other PDE situations like in \cite{Eymard2000, Grossmann2005, Moukalled2016}. During the %
last two decades, it has been applied in several engineering branches such as fluid mechanics, heat and mass transfer, %
and petroleum engineering. In the present work, the finite volume technique is applied to the problem \eqref{SemiLineaModl} %
on a bi-dimensional domain. As a consequence, a nonlinear system is obtained and directly solved on Matlab to produce %
single-peaked and multi-peaked discrete solutions.

\section{Application of the finite volume method}

\begin{figure}[]
  \centering
\includegraphics[scale=1]{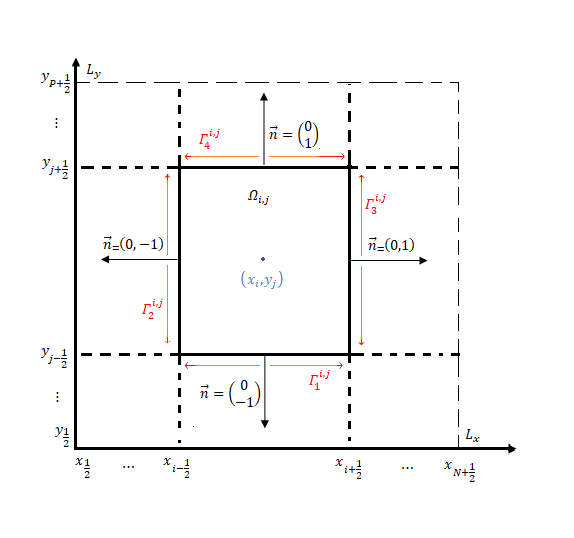} 
\caption{ The control volume $\Omega _{i,j}$ in the domain $\Omega$. }\label{discritisation}
\end{figure}

Consider the problem \eqref{SemiLineaModl} on the rectangular domain $
\Omega =\left] L_{x_1},L_{x_2}\right[ \times \left] L_{y_1},L_{y_2}\right[ $. In addition, the boundary is set as $\partial \Omega =\Gamma _{1}\cup \Gamma _{2}\cup \Gamma _{3}\cup \Gamma
_{4}$, with:%
\begin{equation}\label{boundary}
\Gamma _{1}=\left[ L_{x_1},L_{x2}\right] \times \left\{ L_{y_1}\right\}, \quad \Gamma _{2}=\left\{ L_{x_1}\right\} \times \left[ L_{y_1},L_{y_2}\right], \quad \Gamma _{3}=\left\{ L_{x_2}\right\} \times \left[ L_{y_1},L_{y}\right], \quad \Gamma _{4}=\left[ L_{x_1},L_{x_2}\right] \times \left\{ L_{y_2}\right\}.
\end{equation}
$\Omega$ is partitioned into $N\times P$ control volumes $\Omega _{i,j}$ with center points $(x_{i},y_{j})$ ($i=1, \hdots, N$ and $j=1, \hdots, P$). By choosing the midpoints: $x_{i+1/2}=\dfrac{x_i +x_{i+1}}{2}$, $y_{i+1/2}=\dfrac{y_i +y_{i+1}}{2}$, set:
$h_{x}=x_{i+1/2}-x_{i-1/2}$ and $h_{y}=y_{j+1/2}-y_{j-1/2}$ as the steps. \\
Therefore,
\begin{equation}
\Omega _{i,j}=\left] x_{i-\frac{1}{2}},x_{i+\frac{1}{2}}\right[ \times \left]
y_{j-\frac{1}{2}},y_{j+\frac{1}{2}}\right[.
\end{equation}
The boundary of each control volume $\Omega _{i,j}$ is denoted by $\partial \Omega _{i,j}$ with $\partial \Omega _{i,j}=\bigcup_{i=1}^k\Gamma _{k}^{i,j}$, and we set:
\begin{gather}
\Gamma _{1}^{i,j}=\left[ x_{i-\frac{1}{2}},x_{i+\frac{1}{2}}\right] \times
\left\{ y_{j-\frac{1}{2}}\right\}, \quad \Gamma
_{2}^{i,j}=\left\{ x_{i-\frac{1}{2}}\right\} \times \left[ y_{j-\frac{1}{2}%
},y_{j+\frac{1}{2}}\right],  \notag \\
\Gamma _{3}^{i,j}=\left\{ x_{i+\frac{1}{2}}\right\} \times \left[ y_{j-\frac{%
1}{2}},y_{j+\frac{1}{2}}\right], \quad \Gamma _{4}^{i,j}=\left[
x_{i-\frac{1}{2}},x_{i+\frac{1}{2}}\right] \times \left\{ y_{j+\frac{1}{2}%
}\right\}.
\end{gather}
A representative control volume $\Omega _{i,j}$ of the domain discretization is illustrated in Figure \ref{discritisation}. \\

The discrete solution is assumed to be constant in each control volume $\Omega _{i,j}$ and equal to an approximate %
value $u_{ij}$ of the average $u(x_{i},y_{j})$ in the control volume $\Omega _{i,j}$. \\
Finite volume discretization starts from integrating the PDE of the problem \eqref{SemiLineaModl} on each control volume. In so doing, we get
\begin{equation}
-d\int_{\Omega _{i,j}}\Delta u(x,y)\text{ }dxdy+\int_{\Omega _{i,j}}u(x,y)%
\text{ }dxdy=\int_{\Omega _{i,j}}\left[ u(x,y)\right] ^{q}dxdy, 
\end{equation}%
so that the divergence formula yields
\begin{equation}
-d\int_{\partial \Omega _{i,j}}\nabla u\cdot \overrightarrow{n}\text{ }%
ds+h_{x}h_{y}u_{i,j}=h_{x}h_{y}\left( u_{i,j}\right) ^{q}.
\end{equation}%
Therefore, we obtain
\begin{equation}\label{discretizedeq}
\small -d\int_{\Gamma _{1}^{i,j}}\nabla u\cdot \overrightarrow{n}\text{ }%
ds-d\int_{\Gamma _{2}^{i,j}}\nabla u\cdot \overrightarrow{n}\text{ }%
ds-d\int_{\Gamma _{3}^{i,j}}\nabla u\cdot \overrightarrow{n}\text{ }%
ds-d\int_{\Gamma _{4}^{i,j}}\nabla u\cdot \overrightarrow{n}\text{ }%
ds+h_{x}h_{y}u_{i,j}=h_{x}h_{y}\left( u_{i,j}\right) ^{q}.
\end{equation}

Now, let us proceed to fluxes calculation $\int_{\Gamma_{k}^{i,j}}\nabla u.\overrightarrow{n}$ $ds$ for %
$k=1,2,3,4$. For $k=1$, we have $\overrightarrow{n}=\dbinom{0}{-1}$, so that:
\begin{equation}
\begin{aligned}
\int_{\Gamma _{1}^{i,j}}\nabla u\cdot \overrightarrow{n}\text{ }ds
=&-\int_{\Gamma _{1}^{i,j}}\frac{\partial u}{\partial y}\text{ \ }ds\text{ }
\\
=&-\int_{x_{i-\frac{1}{2}}}^{x_{i+\frac{1}{2}}}\frac{\partial u}{\partial y}%
\left( x,y_{j-\frac{1}{2}}\right) \text{\ }dx.
\end{aligned}
\end{equation}
By selecting the average value of $\frac{\partial u}{\partial y} \left( x,y_{j-\frac{1}{2}}\right) $ on the %
segment $\left[ x_{i-\frac{1}{2}},x_{i+\frac{1}{2}}\right] $ as being $\frac{\partial u}{\partial y}%
\left( x_{i},y_{j-\frac{1}{2}}\right) $, we can find: %
\begin{equation}
\int_{\Gamma _{1}^{i,j}}\nabla u\cdot \overrightarrow{n}\text{ }ds=-h_{x}%
\frac{\partial u}{\partial y}\left( x_{i},y_{j-\frac{1}{2}}\right).
\end{equation}
On the other hand, an approximation of $\frac{\partial u}{\partial y}\left(
x_{i},y_{j-\frac{1}{2}}\right)$ can be given by:

\begin{equation}
\frac{\partial u}{\partial y}\left( x_{i},y_{j-\frac{1}{2}}\right) \cong 
\frac{u_{i,j}-u_{i,j-1}}{h_{y}}.
\end{equation}
Hence,
\begin{equation}\label{flux1}
\int_{\Gamma _{1}^{i,j}}\nabla u\cdot \overrightarrow{n}ds=\frac{h_{x}}{h_{y}%
}\left( u_{i,j-1}-u_{i,j}\right) . 
\end{equation}
Similar steps can be repeated for evaluating the other integrals in \eqref{discretizedeq}. We successively get:

\noindent for $k=2$ with $\overrightarrow{n}=\dbinom{-1}{0}$:

\begin{equation}\label{flux2}
\int_{\Gamma _{2}^{i,j}}\nabla u.\overrightarrow{n}\text{ }ds=\frac{h_{y}}{%
h_{x}}(u_{i-1,j}-u_{i,j}),
\end{equation}
for $k=3$ with $\overrightarrow{n}=\dbinom{1}{0}$:

\begin{equation}\label{flux3}
\int_{\Gamma _{3}^{i,j}}\nabla u.\overrightarrow{n}ds=\frac{h_{y}}{h_{x}}%
(u_{i+1,j}-u_{i,j}), 
\end{equation}
for $k=4$ with $\overrightarrow{n}=\dbinom{0}{1}$:

\begin{equation}\label{flux4}
\int_{\Gamma _{4}^{i,j}}\nabla u.\overrightarrow{n}ds=\frac{h_{x}}{h_{y}}%
\left( u_{i,j+1}-u_{i,j}\right). 
\end{equation}%
The substitution of \eqref{flux1}-\eqref{flux4} into \eqref{discretizedeq} yields for all $i=1, \hdots, N$ et $j=1,\hdots,P$:
\begin{equation} \label{eqafterFluxes}
-d\left[ \frac{h_{x}}{h_{y}}\left( u_{i,j-1}-u_{i,j}\right) +\frac{h_{y}}{%
h_{x}}\left( u_{i-1,j}-u_{i,j}\right) +\frac{h_{y}}{h_{x}}\left(
u_{i+1,j}-u_{i,j}\right) +\frac{h_{x}}{h_{y}}\left( u_{i,j+1}-u_{i,j}\right) %
\right] +h_{x}h_{y}\left[ u_{i,j}-\left( u_{i,j}\right) ^{q}%
\right] =0.
\end{equation}

Next, depending on the control volume $\Omega _{i,j}$, this equation takes different forms because boundary processing %
requires special attention. First, it is important to note that this equation is valid for any control volume whose boundary %
does not meet $\partial \Omega $. 
This concerns control volumes $\Omega_{i,j}$ with $i=2,\hdots,N-1$ and $j=2,\hdots, P-1$. For instance, equation %
\eqref{eqafterFluxes} can display a new form for the control volume $\Omega _{1,1}$ using the boundary condition. %
Therefore, as the Neumann condition is homogeneous on the boundary, it follows that %$\nabla u\cdot \overrightarrow{n}=0$ in $\Gamma_{1}^{1,1}$\ et $\Gamma _{2}^{1,1}$. This gives respectively %
$\int_{\Gamma_{1}^{1,1}}\nabla u\cdot \overrightarrow{n} \, ds=\int_{\Gamma_{2}^{1,1}}\nabla u\cdot \overrightarrow{n} \, =0$,  as $u_{1,0}=u_{1,1}$ and $u_{0,1}=u_{1,1}$. In this case, equation \eqref{eqafterFluxes} can be rewritten in the form:
\begin{equation}
-d\left[ \frac{h_{y}}{h_{x}}\left( u_{2,1}-u_{1,1}\right) +\frac{h_{x}}{h_{y}%
}\left( u_{1,2}-u_{1,1}\right) \right] +h_{x}h_{y}\left[ u_{1,1}-\left(
u_{1,1}\right) ^{q}\right] =0\text{ }  
\end{equation}
Following the same calculation path, we can obtain the corresponding equation for each control volume neighboring %
the boundary of the domain. Finally, the discretization by finite volumes is summarized by the following %
system of nonlinear equations:
\begin{equation} \label{firstchesys}
\left\{ 
\begin{array}{c}
{\small j=1}\text{ \ \ \ \ \ \ \ \ \ \ \ \ \ \ \ \ \ \ \ \ \ \ \ \ \ \ \ \ \
\ \ \ \ \ \ \ \ \ \ \ \ \ \ \ \ \ \ \ \ \ \ \ \ \ \ \ \ \ \ \ \ \ \ \ \ \ \
\ \ \ \ \ \ \ \ \ \ \ \ \ \ \ \ \ \ \ } \\ 
\left\{ 
\begin{array}{c}
{\small i=1}\text{ \ \ \ \ \ \ \ \ \ \ \ \ \ \ \ \ \ \ \ \ \ \ \ \ \ \ \ \ \
\ \ \ \ \ \ \ \ \ \ \ \ \ \ \ \ \ \ \ \ \ \ \ \ \ \ \ \ \ \ \ \ \ \ \ \ \ \
\ \ \ \ \ \ \ \ \ \ \ \ \ \ \ \ \ \ \ \ \ \ } \\ 
\left[ h_{x}h_{y}+d\left( \frac{h_{x}}{h_{y}}+\frac{h_{y}}{h_{x}}\right) %
\right] {\small u}_{1,1}{\small -d}\frac{h_{y}}{h_{x}}{\small u}_{2,1}%
{\small -d}\frac{h_{x}}{h_{y}}{\small u}_{1,2}{\small -h}_{x}{\small h}%
_{y}\left( u_{1,1}\right) ^{q}{\small =0}\text{ \ \ \ \ \ \ \ \ \ \ \ \ \ \
\ \ \ \ \ \ \ \ \ \ \ \ \ \ \ \ \ \ \ \ \ \ \ \ \ \ \ \ \ \ \ \ \ \ \ \ \ \
\ \ \ \ \ \ \ \ \ \ \ \ \ \ \ \ \ \ \ \ \ \ \ \ \ \ \ \ \ \ \ \ \ \ \ \ \ \ }
\\ 
{\small i=}\overline{\text{\ }2,N-1}\text{ \ \ \ \ \ \ \ \ \ \ \ \ \ \ \ \ \
\ \ \ \ \ \ \ \ \ \ \ \ \ \ \ \ \ \ \ \ \ \ \ \ \ \ \ \ \ \ \ \ \ \ \ \ \ \
\ \ \ \ \ \ \ \ \ \ \ \ \ \ \ \ \ \ \ \ \ \ \ \ \ \ \ \ \ \ \ \ } \\ 
{\small -d}\frac{h_{y}}{h_{x}}{\small u}_{i-1,1}{\small +}\left[
h_{x}h_{y}+d\left( \frac{h_{x}}{h_{y}}+2\frac{h_{y}}{h_{x}}\right) \right] 
{\small u}_{i,1}{\small -d}\frac{h_{y}}{h_{x}}{\small u}_{i+1,1}{\small -d}%
\frac{h_{x}}{h_{y}}{\small u}_{i,2}{\small -h}_{x}{\small h}_{y}\left(
u_{i,1}\right) ^{q}{\small =0}\text{ \ \ \ \ \ \ \ \ \ \ \ \ \ \ \ \ \ \ \ \
\ \ \ \ \ \ \ \ \ \ \ \ \ \ \ \ \ \ \ \ \ \ \ \ \ \ \ \ \ \ \ \ \ \ \ \ \ \
\ \ \ \ \ \ \ \ \ \ \ \ \ \ \ \ } \\ 
{\small i=N}\text{\ \ \ \ \ \ \ \ \ \ \ \ \ \ \ \ \ \ \ \ \ \ \ \ \ \ \ \ \
\ \ \ \ \ \ \ \ \ \ \ \ \ \ \ \ \ \ \ \ \ \ \ \ \ \ \ \ \ \ \ \ \ \ \ \ \ \
\ \ \ \ \ \ \ \ \ \ \ \ \ \ \ \ \ \ \ \ \ \ } \\ 
{\small -d}\frac{h_{y}}{h_{x}}{\small u}_{N-1,1}{\small +}\left[
h_{x}h_{y}+d\left( \frac{h_{x}}{h_{y}}+\frac{h_{y}}{h_{x}}\right) \right] 
{\small u}_{N,1}{\small -d}\frac{h_{x}}{h_{y}}{\small u}_{N,2}{\small -h}_{x}%
{\small h}_{y}\left( u_{N,1}\right) ^{q}{\small =0}\text{ \ \ \ \ \ \ \ \ \
\ \ \ \ \ \ \ \ \ \ \ \ \ \ \ \ \ \ \ \ \ \ \ \ \ \ \ \ \ \ \ \ \ \ \ \ \ \
\ \ \ \ \ \ \ \ \ \ \ \ \ \ \ \ \ \ \ \ \ \ \ \ \ \ \ \ \ \ \ \ \ \ \ \ \ \ }%
\end{array}%
\right. \\ 
{\small j=}\overline{2,P-1}\text{ \ \ \ \ \ \ \ \ \ \ \ \ \ \ \ \ \ \ \ \ \
\ \ \ \ \ \ \ \ \ \ \ \ \ \ \ \ \ \ \ \ \ \ \ \ \ \ \ \ \ \ \ \ \ \ \ \ \ \
\ \ \ \ \ \ \ \ \ \ \ \ \ \ \ \ \ \ \ \ \ \ \ \ } \\ 
\left\{ 
\begin{array}{c}
{\small i=1}\text{\ \ \ \ \ \ \ \ \ \ \ \ \ \ \ \ \ \ \ \ \ \ \ \ \ \ \ \ \
\ \ \ \ \ \ \ \ \ \ \ \ \ \ \ \ \ \ \ \ \ \ \ \ \ \ \ \ \ \ \ \ \ \ \ \ \ \
\ \ \ \ \ \ \ \ \ \ \ \ \ \ \ \ \ \ \ \ \ \ } \\ 
{\small -d}\frac{h_{x}}{h_{y}}{\small u}_{1,j-1}{\small +}\left[
h_{x}h_{y}+d\left( 2\frac{h_{x}}{h_{y}}+\frac{h_{y}}{h_{x}}\right) \right] 
{\small u}_{1,j}{\small -d}\frac{h_{y}}{h_{x}}{\small u}_{2,j}{\small -d}%
\frac{h_{x}}{h_{y}}{\small u}_{1,j+1}{\small -h}_{x}{\small h}_{y}\left(
u_{1,j}\right) ^{q}{\small =0}\text{ \ \ \ \ \ \ \ \ \ \ \ \ \ \ \ \ \ \ \ \
\ \ \ \ \ \ \ \ \ \ \ \ \ \ \ \ \ \ \ \ \ \ \ \ \ \ \ \ \ \ \ \ \ \ \ \ \ \
\ \ \ \ \ \ \ \ \ \ \ \ } \\ 
{\small i=}\overline{\text{\ }2,N-1}\text{ \ \ \ \ \ \ \ \ \ \ \ \ \ \ \ \ \
\ \ \ \ \ \ \ \ \ \ \ \ \ \ \ \ \ \ \ \ \ \ \ \ \ \ \ \ \ \ \ \ \ \ \ \ \ \
\ \ \ \ \ \ \ \ \ \ \ \ \ \ \ \ \ \ \ \ \ \ \ \ \ \ \ \ \ \ \ \ } \\ 
{\small -d}\frac{h_{x}}{h_{y}}{\small u}_{i,j-1}{\small -d}\frac{h_{y}}{h_{x}%
}{\small u}_{i-1,j}{\small +}\left[ h_{x}h_{y}+2d\left( \frac{h_{x}}{h_{y}}+%
\frac{h_{y}}{h_{x}}\right) \right] {\small u}_{i,j}{\small -d}\frac{h_{y}}{%
h_{x}}{\small u}_{i+1,j}-\text{ \ \ \ \ \ \ \ \ \ \ \ \ \ \ \ \ \ \ \ \ \ \
\ \ \ \ \ \ \ \ \ \ \ \ \ \ \ \ \ \ \ \ \ \ \ \ \ \ \ \ \ \ \ \ \ \ \ \ \ \
\ \ \ \ \ \ \ \ \ \ \ \ \ \ \ \ \ \ \ \ \ \ \ \ \ \ \ \ \ } \\ 
{\small d}\frac{h_{x}}{h_{y}}{\small u}_{i,j+1}{\small -h}_{x}{\small h}%
_{y}\left( u_{i,j}\right) ^{q}{\small =0}\text{\ \ \ \ \ \ \ \ \ \ } \\ 
{\small i=N}\text{ \ \ \ \ \ \ \ \ \ \ \ \ \ \ \ \ \ \ \ \ \ \ \ \ \ \ \ \ \
\ \ \ \ \ \ \ \ \ \ \ \ \ \ \ \ \ \ \ \ \ \ \ \ \ \ \ \ \ \ \ \ \ \ \ \ \ \
\ \ \ \ \ \ \ \ \ \ \ \ \ } \\ 
{\small -d}\frac{h_{x}}{h_{y}}{\small u}_{N,j-1}{\small -d}\frac{h_{y}}{h_{x}%
}{\small u}_{N-1,j}{\small +}\left[ h_{x}h_{y}+d\left( 2\frac{h_{x}}{h_{y}}+%
\frac{h_{y}}{h_{x}}\right) \right] {\small u}_{N,j}{\small -d}\frac{h_{x}}{%
h_{y}}{\small u}_{N,j+1}{\small -}\text{ \ \ \ \ \ \ \ \ \ \ \ \ \ \ \ \ \ \
\ \ \ \ \ \ \ \ \ \ \ \ \ \ \ \ \ \ \ \ \ \ \ \ \ \ \ \ \ \ \ \ \ \ \ \ \ \
\ \ \ \ \ \ \ \ \ \ \ \ \ \ \ \ \ \ \ \ \ \ \ \ \ \ \ \ \ \ \ \ } \\ 
\text{ \ \ \ \ \ \ \ \ \ \ \ \ \ \ \ \ \ \ \ \ \ \ \ \ \ \ \ \ \ \ \ \ \ \ \
\ \ \ \ \ \ \ \ \ \ \ \ \ \ \ \ \ \ \ \ \ \ \ \ \ \ \ \ \ \ \ \ \ \ \ \ \ \
\ \ \ \ \ }{\small h}_{x}{\small h}_{y}\left( u_{N,j}\right) ^{q}{\small =0}%
\text{ \ \ \ \ \ \ \ \ \ \ \ \ \ \ \ \ \ \ \ \ \ \ \ \ \ \ \ \ \ \ \ \ \ \ \
\ \ \ \ \ \ \ \ \ \ \ \ \ \ \ \ \ \ \ \ \ \ \ \ \ \ \ \ \ \ \ \ \ \ \ \ \ \ }
\\ 
\text{\ \ \ \ \ \ \ \ }%
\end{array}%
\right. \\ 
{\small j=P}\text{ \ \ \ \ \ \ \ \ \ \ \ \ \ \ \ \ \ \ \ \ \ \ \ \ \ \ \ \ \
\ \ \ \ \ \ \ \ \ \ \ \ \ \ \ \ \ \ \ \ \ \ \ \ \ \ \ \ \ \ \ \ \ \ \ \ \ \
\ \ \ \ \ \ \ } \\ 
\left\{ 
\begin{array}{c}
{\small i=1}\text{ \ \ \ \ \ \ \ \ \ \ \ \ \ \ \ \ \ \ \ \ \ \ \ \ \ \ \ \ \
\ \ \ \ \ \ \ \ \ \ \ \ \ \ \ \ \ \ \ \ \ \ \ \ \ \ \ \ \ \ \ \ \ \ \ \ \ \
\ \ \ \ \ \ \ \ \ \ \ } \\ 
{\small -d}\frac{h_{x}}{h_{y}}{\small u}_{1,P-1}{\small +}\left[
h_{x}h_{y}+d\left( \frac{h_{x}}{h_{y}}+\frac{h_{y}}{h_{x}}\right) \right] 
{\small u}_{1,P}{\small -d}\frac{h_{y}}{h_{x}}{\small u}_{2,P}{\small -h}_{x}%
{\small h}_{y}\left( u_{1,P}\right) ^{q}{\small =0}\text{ \ \ \ \ \ \ \ \ \
\ \ \ \ \ \ \ \ \ \ \ \ \ \ \ \ \ \ \ \ \ \ \ \ \ \ \ \ \ \ \ \ \ \ \ \ \ \
\ \ \ \ \ \ \ \ \ \ \ \ \ \ \ \ \ \ \ \ \ \ \ \ \ \ \ \ \ \ \ \ \ \ \ \ \ \ }
\\ 
{\small i=}\overline{\text{\ }2,N-1}\text{\ \ \ \ \ \ \ \ \ \ \ \ \ \ \ \ \
\ \ \ \ \ \ \ \ \ \ \ \ \ \ \ \ \ \ \ \ \ \ \ \ \ \ \ \ \ \ \ \ \ \ \ \ \ \
\ \ \ \ \ \ \ \ \ \ \ \ \ \ \ \ \ \ \ \ \ \ } \\ 
{\small -d}\frac{h_{x}}{h_{y}}{\small u}_{i,P-1}{\small -d}\frac{h_{y}}{h_{x}%
}{\small u}_{i-1,P}{\small +}\left[ h_{x}h_{y}+d\left( 2\frac{h_{y}}{h_{x}}+%
\frac{h_{x}}{h_{y}}\right) \right] {\small u}_{i,p}{\small -d}\frac{h_{y}}{%
h_{x}}{\small u}_{i+1,P}{\small -}\text{ \ \ \ \ \ \ \ \ \ \ \ \ \ \ \ \ \ \
\ \ \ \ \ \ \ \ \ \ \ \ \ \ \ \ \ \ \ \ \ \ \ \ \ \ \ \ \ \ \ \ \ \ \ \ \ \
\ \ \ \ \ \ \ \ \ \ \ \ \ \ \ \ \ \ \ \ \ \ \ \ \ \ \ \ \ \ \ \ \ } \\ 
{\small h}_{x}{\small h}_{y}\left( u_{i,P}\right) ^{q}=0\text{\ \ \ \ } \\ 
{\small i=N}\text{ \ \ \ \ \ \ \ \ \ \ \ \ \ \ \ \ \ \ \ \ \ \ \ \ \ \ \ \ \
\ \ \ \ \ \ \ \ \ \ \ \ \ \ \ \ \ \ \ \ \ \ \ \ \ \ \ \ \ \ \ \ \ \ \ \ \ \
\ \ \ \ \ \ \ \ \ \ } \\ 
{\small -d}\frac{h_{x}}{h_{y}}{\small u}_{N,P-1}{\small -d}\frac{h_{y}}{h_{x}%
}{\small u}_{N-1,P}{\small +}\left[ h_{x}h_{y}+d\left( \frac{h_{y}}{h_{x}}+%
\frac{h_{x}}{h_{y}}\right) \right] {\small u}_{N,p}{\small -h}_{x}{\small h}%
_{y}\left( u_{N,P}\right) ^{q}{\small =0}\text{ \ \ \ \ \ \ \ \ \ \ \ \ \ \
\ \ \ \ \ \ \ \ \ \ \ \ \ \ \ \ \ \ \ \ \ \ \ \ \ \ \ \ \ \ \ \ \ \ \ \ \ \
\ \ \ \ \ \ \ \ \ \ \ \ \ \ \ \ \ \ \ \ \ \ \ \ \ \ \ \ \ }%
\end{array}%
\right.%
\end{array}%
\right. 
\end{equation}

\section{Numerical test}
\begin{figure}[]
\begin{center}
\includegraphics[scale=0.6]{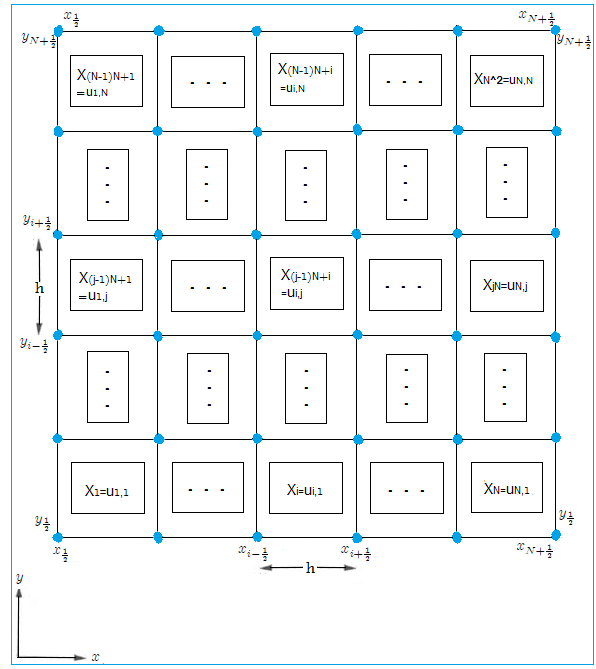}
\end{center}
\caption{Notation illustration.}\label{notation}
\end{figure}

In this section, we assume that $L_{x_1}=L{y_1}$ and $L_{x_2}=L_{y_2}$ and we suppose a uniform mesh %
by setting: $h_{x}=h_{y}=h $. As a consequence, we get $N=P$. Introducing a new notation of $u_{i,j}$ %
(for $i=\overline{1,N}$ and  $j=\overline{1,N}$) of the system \eqref{SemiLineaModl} by %
$X_{S}$ (for $S=\overline{1,N^{2}}$) such that single-index numbering is performed conventionally from  %
left to right and from bottom to top. We also take $s=(j-1)N+i$ for $j=\overline{1,N}$ and $i=\overline{1,N}$ %
as in Figure~\ref{notation}. \\

For the sake of simplification, we set $a=h^{2}+2d$, $b=h^{2}+3d$, and $c=h^{2}+4d$ in the system \eqref{firstchesys} to get the  following equivalent nonlinear system:
\\
\\
\begin{equation} \label{secondchesys}
\left\{ 
\begin{array}{c}
j=1\text{ \ \ \ \ \ \ \ \ \ \ \ \ \ \ \ \ \ \ \ \ \ \ \ \ \ \ \ \ \ \ \ \ \
\ \ \ \ \ \ \ \ \ \ \ \ \ \ \ \ \ \ \ \ \ \ \ \ \ \ \ \ \ \ \ \ \ \ \ \ \ }
\\ 
\left\{ 
\begin{array}{c}
i=1\text{ \ \ \ \ \ \ \ \ \ \ \ \ \ \ \ \ \ \ \ \ \ \ \ \ \ \ \ \ \ \ \ \ \
\ \ \ \ \ \ \ \ \ \ \ \ \ \ \ \ \ \ \ \ \ \ \ \ \ \ \ \ \ \ \ \ \ \ \ \ \ \
\ \ } \\ 
\left( a-h^{2}X_{1}^{q-1}\right) X_{1}-dX_{2}-dX_{N+1}=0,\text{ \ \ \ \ \ \ \
\ \ \ \ \ \ \ \ \ \ \ \ \ \ \ \ \ \ \ \ \ \ \ \ \ \ \ \ \ \ \ \ \ \ \ \ \ \
\ \ \ \ \ \ \ \ \ \ \ \ \ \ \ \ \ \ \ \ \ \ \ \ \ \ \ \ \ \ \ \ \ \ \ \ \ \
\ \ \ \ \ \ \ \ \ \ \ \ \ \ \ \ \ \ \ \ \ \ \ } \\ 
i=\overline{2,N-1}\text{ \ \ \ \ \ \ \ \ \ \ \ \ \ \ \ \ \ \ \ \ \ \ \ \ \ \
\ \ \ \ \ \ \ \ \ \ \ \ \ \ \ \ \ \ \ \ \ \ \ \ \ \ \ \ \ \ \ \ \ \ \ \ \ \
\ \ \ \ \ \ \ \ } \\ 
-dX_{i-1}+\left( b-h^{2}X_{i}^{q-1}\right) X_{i}-dX_{i+1}-dX_{i+N}=0,\text{ \
\ \ \ \ \ \ \ \ \ \ \ \ \ \ \ \ \ \ \ \ \ \ \ \ \ \ \ \ \ \ \ \ \ \ \ \ \ \
\ \ \ \ \ \ \ \ \ \ \ \ \ \ \ \ \ \ \ \ \ \ \ \ \ \ \ \ \ \ \ \ \ \ \ \ \ \
\ \ \ \ \ \ \ \ \ \ \ \ \ } \\ 
i=N\text{ \ \ \ \ \ \ \ \ \ \ \ \ \ \ \ \ \ \ \ \ \ \ \ \ \ \ \ \ \ \ \ \ \
\ \ \ \ \ \ \ \ \ \ \ \ \ \ \ \ \ \ \ \ \ \ \ \ \ \ \ \ \ \ \ \ \ \ \ \ \ \
\ } \\ 
-dX_{N-1}+\left( a-h^{2}X_{N}^{q-1}\right) X_{N}-dX_{2N}=0,\text{ \ \ \ \ \ \
\ \ \ \ \ \ \ \ \ \ \ \ \ \ \ \ \ \ \ \ \ \ \ \ \ \ \ \ \ \ \ \ \ \ \ \ \ \
\ \ \ \ \ \ \ \ \ \ \ \ \ \ \ \ \ \ \ \ \ \ \ \ \ \ \ \ \ \ \ \ \ \ \ \ \ \
\ \ \ \ \ \ \ \ \ \ \ \ \ \ \ \ \ \ \ \ }%
\end{array}%
\right. \\ 
j=\overline{2,N-1}\text{ \ \ \ \ \ \ \ \ \ \ \ \ \ \ \ \ \ \ \ \ \ \ \ \ \ \
\ \ \ \ \ \ \ \ \ \ \ \ \ \ \ \ \ \ \ \ \ \ \ \ \ \ \ \ \ \ \ \ \ \ \ \ \ \
\ \ \ \ \ \ \ \ \ } \\ 
\left\{ 
\begin{array}{c}
i=1\text{ \ \ \ \ \ \ \ \ \ \ \ \ \ \ \ \ \ \ \ \ \ \ \ \ \ \ \ \ \ \ \ \ \
\ \ \ \ \ \ \ \ \ \ \ \ \ \ \ \ \ \ \ \ \ \ \ \ \ \ \ \ \ \ \ \ \ \ \ \ \ \
\ \ \ \ } \\ 
-dX_{(j-2)N+1}+\left( b-h^{2}X_{(j-1)N+1}^{q-1}\right)
X_{(j-1)N+1}-dX_{(j-1)N+2}-dX_{jN+1}=0,\text{ \ \ \ \ \ \ \ \ \ \ \ \ \ \ \ \
\ \ \ \ \ \ \ \ \ \ \ \ \ \ \ \ \ \ \ \ \ \ \ \ \ \ \ \ \ \ \ \ \ \ \ \ \ \
\ \ } \\ 
i=\overline{2,N-1}\text{ \ \ \ \ \ \ \ \ \ \ \ \ \ \ \ \ \ \ \ \ \ \ \ \ \ \
\ \ \ \ \ \ \ \ \ \ \ \ \ \ \ \ \ \ \ \ \ \ \ \ \ \ \ \ \ \ \ \ \ \ \ \ \ \
\ \ \ \ \ \ \ \ \ \ \ \ \ \ } \\ 
-dX_{(j-2)N+i}-dX_{(j-1)N+i-1}+\left( c-h^{2}X_{(N-1)N+i}^{q-1}\right)
X_{(j-1)N+i}-dX_{(j-1)N+i+1}-dX_{jN+i}=0,\text{\ \ \ \ \ \ \ \ \ \ \ \ \ \ \ \ \ \ \ \ \ 
\ \ \ \ \ \ \ \ \ \ \ \ \ \ } \\ 
i=N\text{ \ \ \ \ \ \ \ \ \ \ \ \ \ \ \ \ \ \ \ \ \ \ \ \ \ \ \ \ \ \ \ \ \
\ \ \ \ \ \ \ \ \ \ \ \ \ \ \ \ \ \ \ \ \ \ \ \ \ \ \ \ \ \ \ \ \ \ \ \ \ \
\ \ \ } \\ 
-dX_{(j-1)N}-dX_{jN-1}+\left( b-h^{2}X_{jN}^{q-1}\right) X_{jN}-dX_{(j+1)N}=0,%
\text{ \ \ \ \ \ \ \ \ \ \ \ \ \ \ \ \ \ \ \ \ \ \ \ \ \ \ \ \ \ \ \ \ \ \ \
\ \ \ \ \ \ \ \ \ \ \ \ \ \ \ \ \ \ \ \ \ \ \ \ \ \ \ \ \ \ \ \ \ \ \ \ \ \
\ \ \ }%
\end{array}%
\right. \\ 
j=N\text{ \ \ \ \ \ \ \ \ \ \ \ \ \ \ \ \ \ \ \ \ \ \ \ \ \ \ \ \ \ \ \ \ \
\ \ \ \ \ \ \ \ \ \ \ \ \ \ \ \ \ \ \ \ \ \ \ \ \ \ \ \ \ \ \ \ \ \ \ \ } \\ 
\left\{ 
\begin{array}{c}
i=1\text{ \ \ \ \ \ \ \ \ \ \ \ \ \ \ \ \ \ \ \ \ \ \ \ \ \ \ \ \ \ \ \ \ \
\ \ \ \ \ \ \ \ \ \ \ \ \ \ \ \ \ \ \ \ \ \ \ \ \ \ \ \ \ \ \ \ \ \ \ \ \ \
\ \ } \\ 
-dX_{(N-2)N+1}+\left( a-h^{2}X_{(N-1)N+1}^{q-1}\right)
X_{(N-1)N+1}-dX_{(N-1)N+2}=0,\text{ \ \ \ \ \ \ \ \ \ \ \ \ \ \ \ \ \ \ \ \ \
\ \ \ \ \ \ \ \ \ \ \ \ \ \ \ \ \ \ \ \ \ \ \ \ \ \ \ \ \ \ \ \ \ \ \ \ \ \
\ \ \ \ \ \ \ \ } \\ 
i=\overline{2,N-1}\text{ \ \ \ \ \ \ \ \ \ \ \ \ \ \ \ \ \ \ \ \ \ \ \ \ \ \
\ \ \ \ \ \ \ \ \ \ \ \ \ \ \ \ \ \ \ \ \ \ \ \ \ \ \ \ \ \ \ \ \ \ \ \ \ \
\ \ \ \ \ \ \ \ } \\ 
-dX_{(N-2)N+i}-dX_{(N-1)N+i-1}+\left( b-h^{2}X_{(N-1)N+i}^{q-1}\right)
X_{(N-1)N+i}-dX_{(N-1)N+i+1}=0,\text{ \ \ \ \ \ \ \ \ \ \ \ \ \ \ \ \ \ \ \ \ \ \ \ \ \ \ \ \ \ \ \ \ \
\ \ \ \ \ \ \ \ \ \ }
\\ 
i=N\text{ \ \ \ \ \ \ \ \ \ \ \ \ \ \ \ \ \ \ \ \ \ \ \ \ \ \ \ \ \ \ \ \ \
\ \ \ \ \ \ \ \ \ \ \ \ \ \ \ \ \ \ \ \ \ \ \ \ \ \ \ \ \ \ \ \ \ \ \ \ \ \
\ } \\ 
-dX_{(N-1)N}-dX_{N^{2}-1}+\left( a-h^{2}X_{N^{2}}^{q-1}\right) X_{N^{2}}=0.%
\text{ \ \ \ \ \ \ \ \ \ \ \ \ \ \ \ \ \ \ \ \ \ \ \ \ \ \ \ \ \ \ \ \ \ \ \
\ \ \ \ \ \ \ \ \ \ \ \ \ \ \ \ \ \ \ \ \ \ \ \ \ \ \ \ \ \ \ \ \ \ \ \ \ \
\ \ \ \ \ \ \ \ \ \ \ \ \ \ \ \ \ \ }%
\end{array}%
\right.%
\end{array}%
\right. 
\end{equation}

\begin{figure}[H]
    \centering
      \begin{subfigure}[b]{0.32\textwidth}
        \centering
          \includegraphics[width=\textwidth]{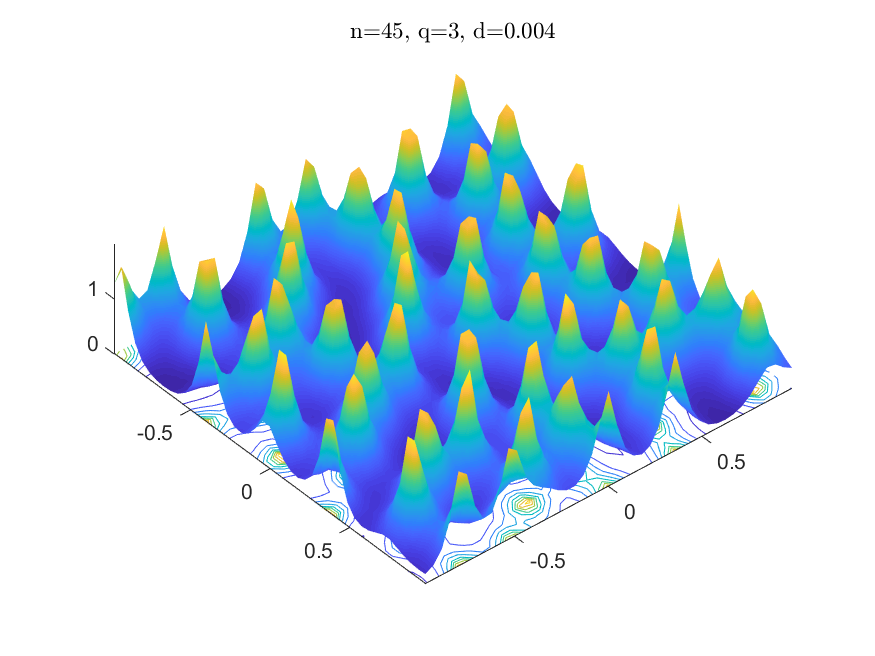}
          \caption{}
          \label{chfig1}
          \end{subfigure}
          \centering
      \begin{subfigure}[b]{0.32\textwidth}
        \centering
          \includegraphics[width=\textwidth]{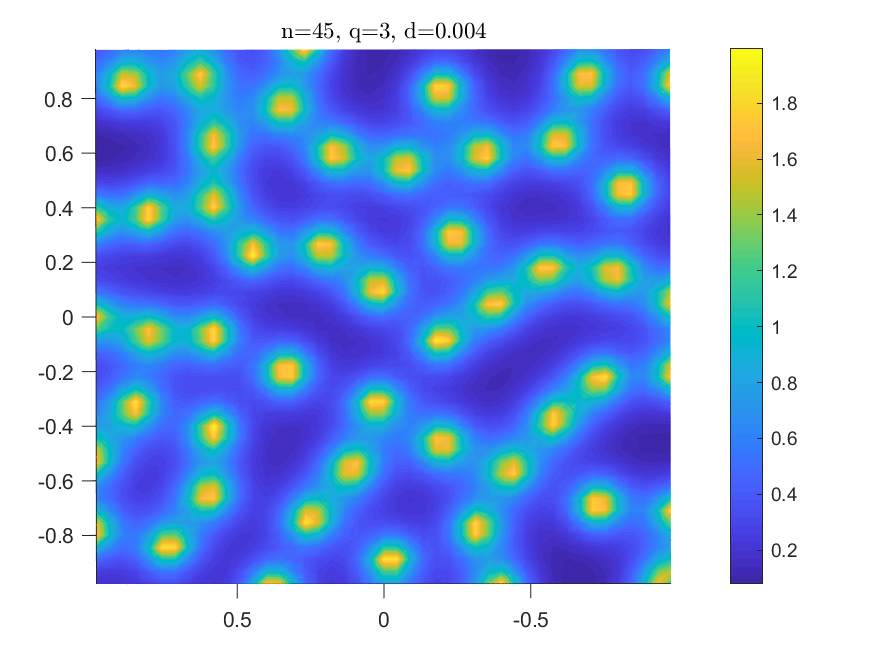}
          \caption{}
          \label{chfig1v}
          \end{subfigure}
           \begin{subfigure}[b]{0.32\textwidth}
        \centering
          \includegraphics[width=\textwidth]{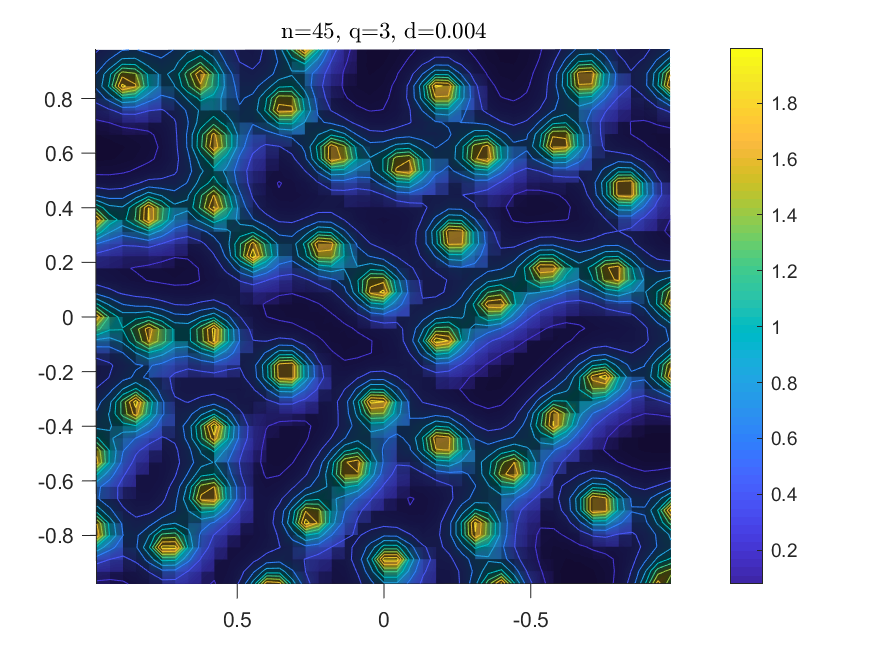}
          \caption{}
          \label{chfig1c}
      \end{subfigure}
      \caption{Numerical simulation of problem \eqref{SemiLineaModl} with $q=3$, $d=0.004$, and $N=45$}\label{cfig1}
\end{figure}

\begin{figure}[H]
    \centering
      \begin{subfigure}[b]{0.32\textwidth}
        \centering
          \includegraphics[width=\textwidth]{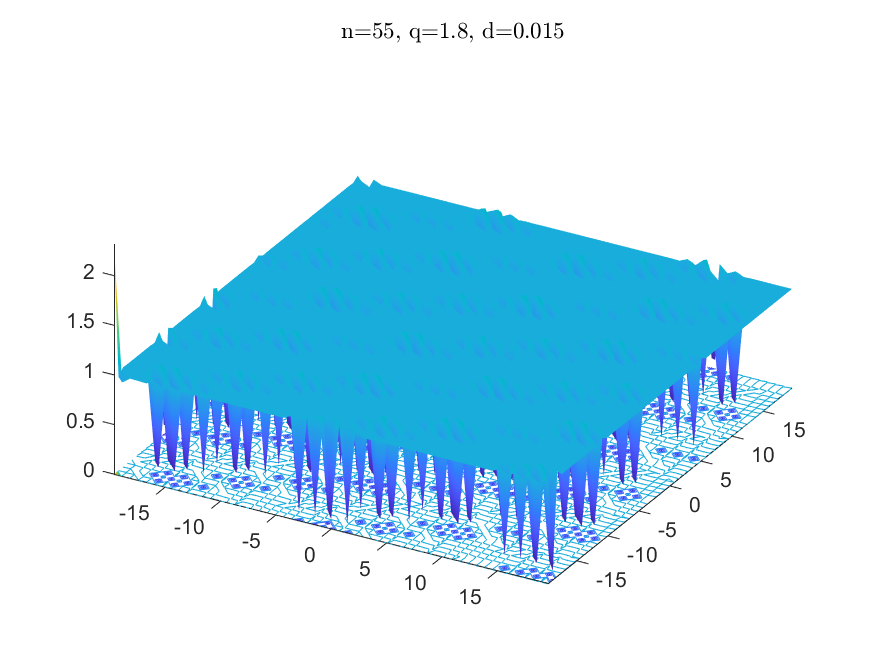}
          \caption{}
          \label{chfig2}
          \end{subfigure}
          \centering
      \begin{subfigure}[b]{0.32\textwidth}
        \centering
          \includegraphics[width=\textwidth]{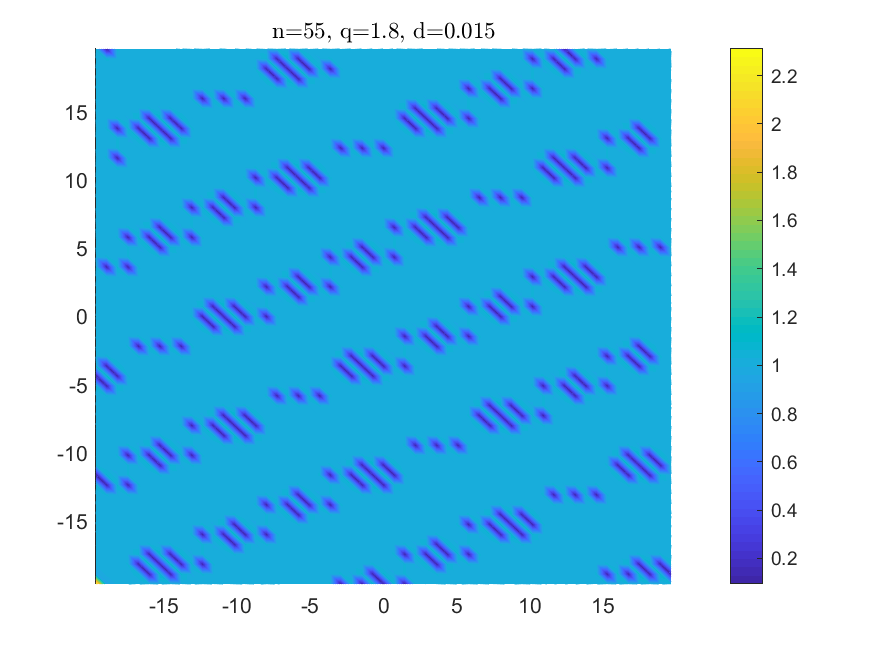}
          \caption{}
          \label{chfig2v}
          \end{subfigure}
           \begin{subfigure}[b]{0.32\textwidth}
        \centering
          \includegraphics[width=\textwidth]{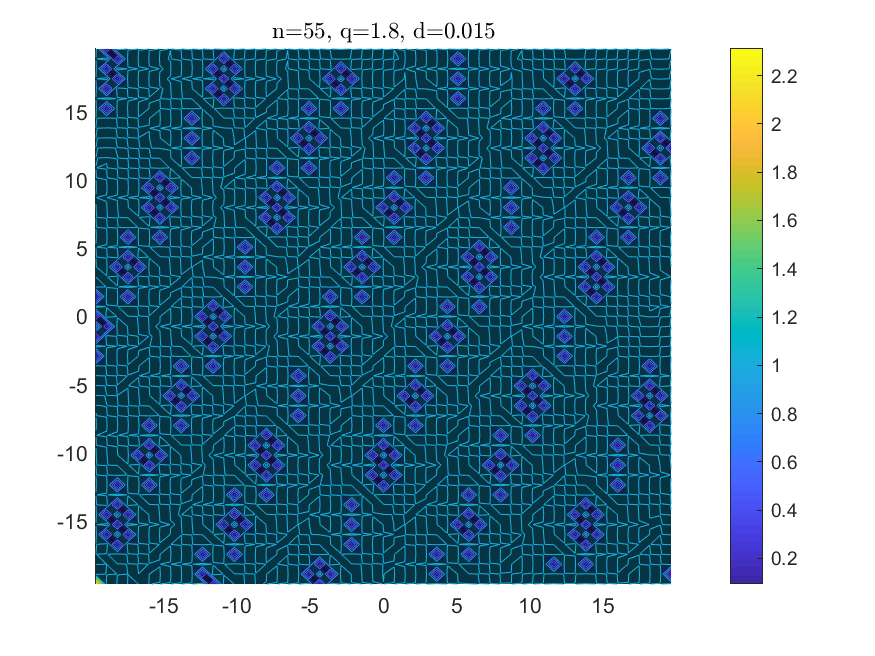}
          \caption{}
          \label{chfig2c}
      \end{subfigure}
      \caption{Numerical simulation of problem \eqref{SemiLineaModl} with $q=1.8$, $d=0.015$, and $N=55$}\label{cfig2}
\end{figure}

\begin{figure}[H]
    \centering
      \begin{subfigure}[b]{0.32\textwidth}
        \centering
          \includegraphics[width=\textwidth]{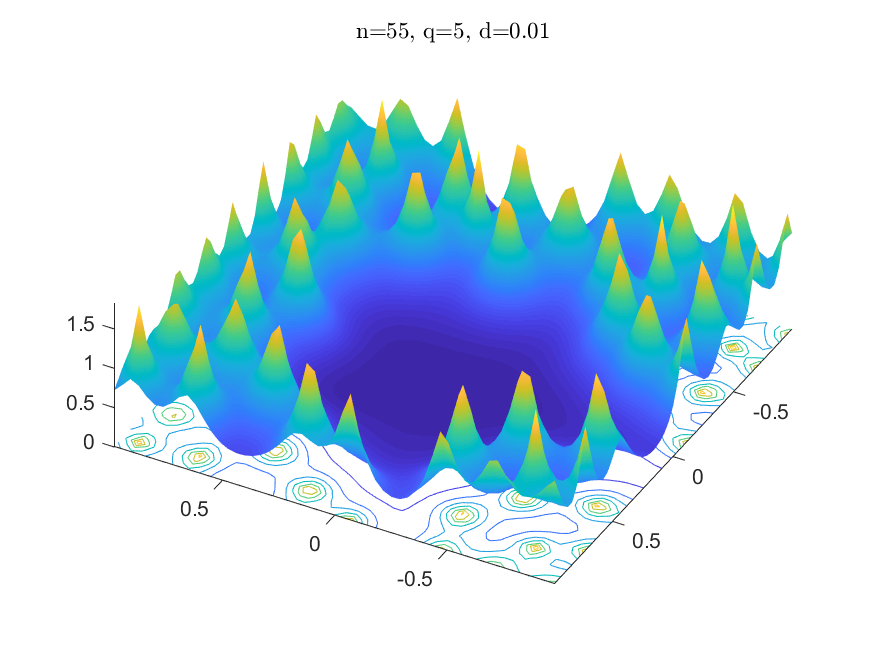}
          \caption{}
          \label{chfig3}
          \end{subfigure}
          \centering
      \begin{subfigure}[b]{0.32\textwidth}
        \centering
          \includegraphics[width=\textwidth]{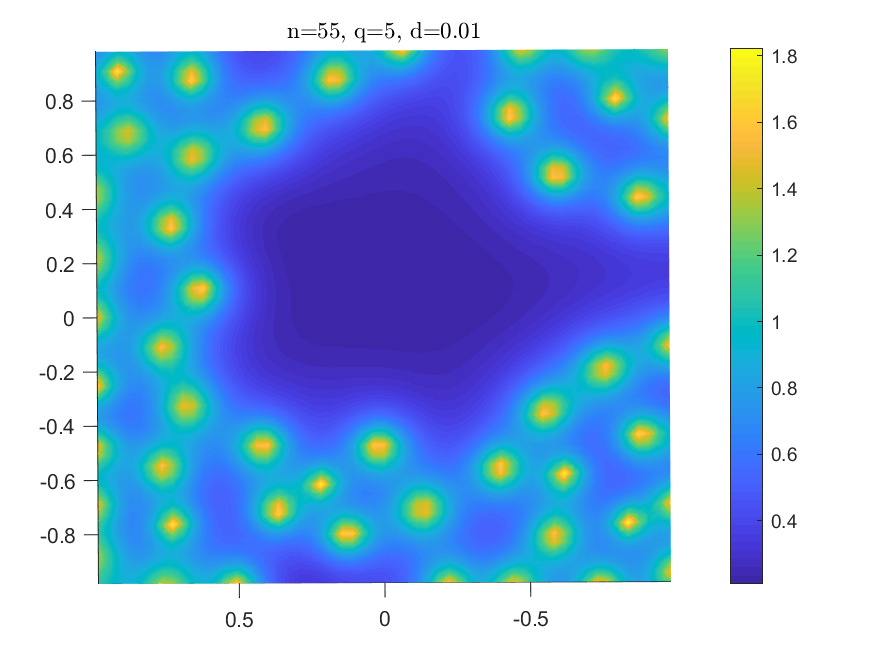}
          \caption{}
          \label{chfig3v}
          \end{subfigure}
           \begin{subfigure}[b]{0.32\textwidth}
        \centering
          \includegraphics[width=\textwidth]{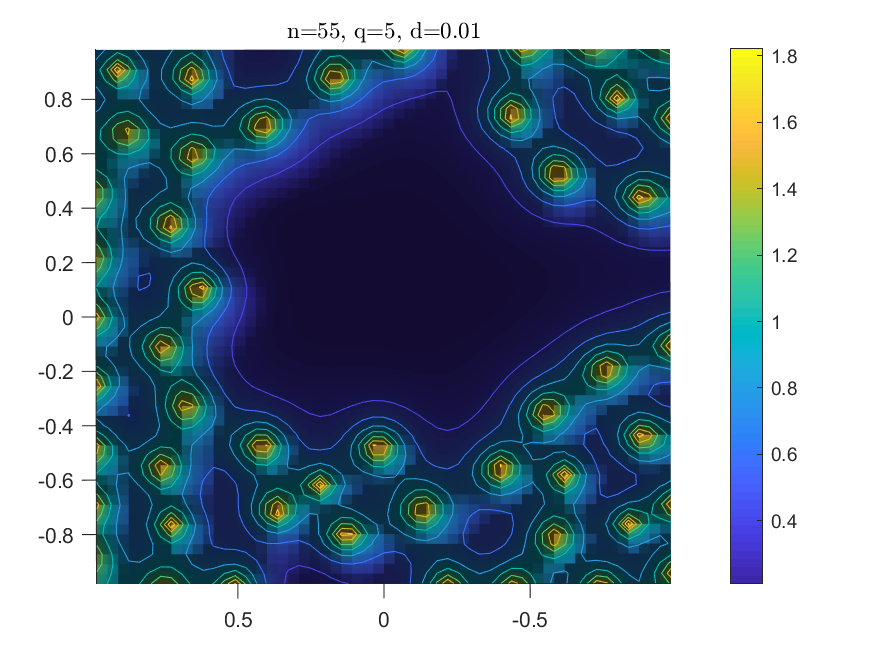}
          \caption{}
          \label{chfig3c}
      \end{subfigure}
      \caption{Numerical simulation of problem \eqref{SemiLineaModl} with $q=5$, $d=0.01$, and $N=55$}\label{cfig3}
\end{figure}

\begin{figure}[H]
    \centering
      \begin{subfigure}[b]{0.32\textwidth}
        \centering
          \includegraphics[width=\textwidth]{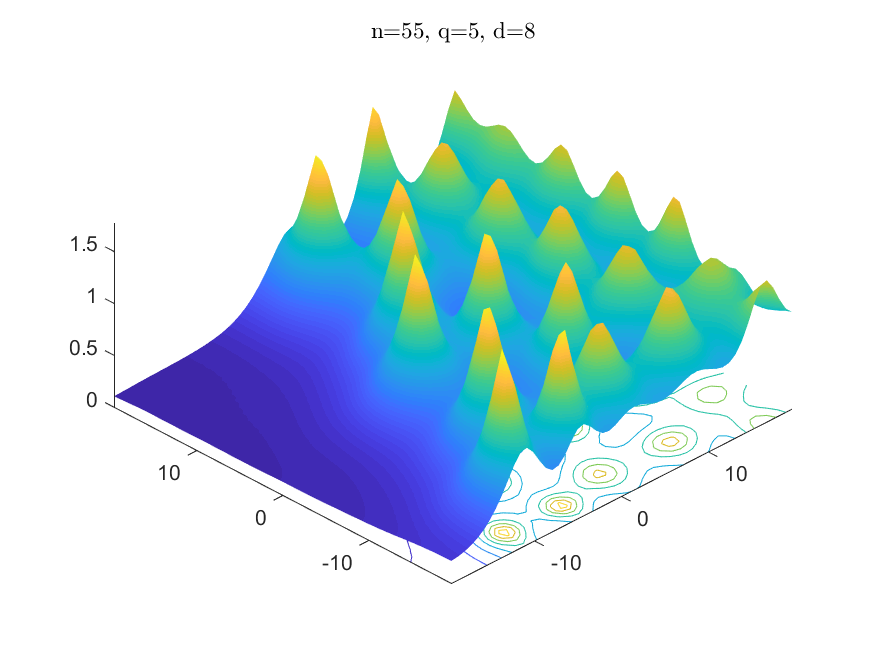}
          \caption{}
          \label{chfig4}
          \end{subfigure}
          \centering
      \begin{subfigure}[b]{0.32\textwidth}
        \centering
          \includegraphics[width=\textwidth]{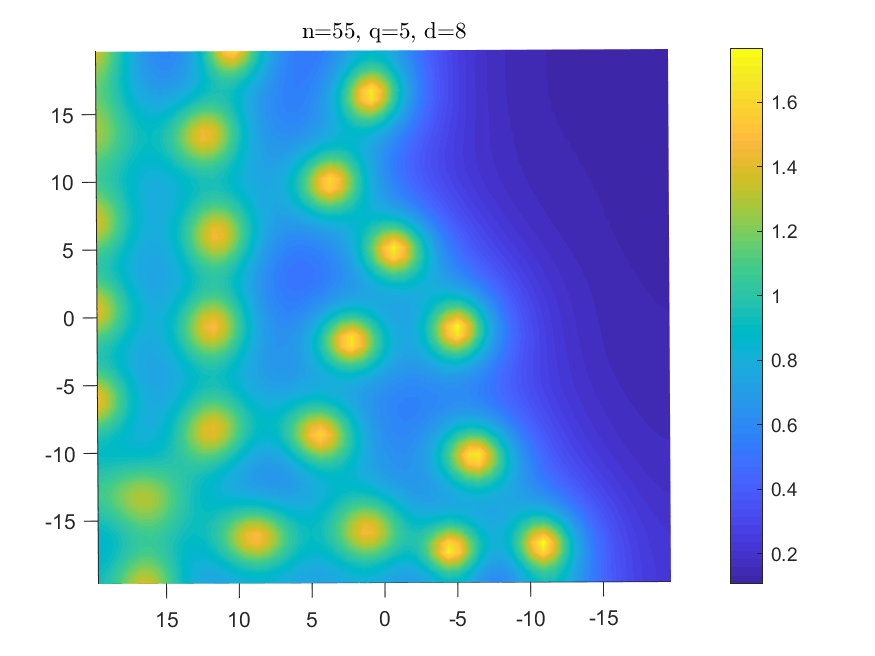}
          \caption{}
          \label{chfig4v}
          \end{subfigure}
           \begin{subfigure}[b]{0.32\textwidth}
        \centering
          \includegraphics[width=\textwidth]{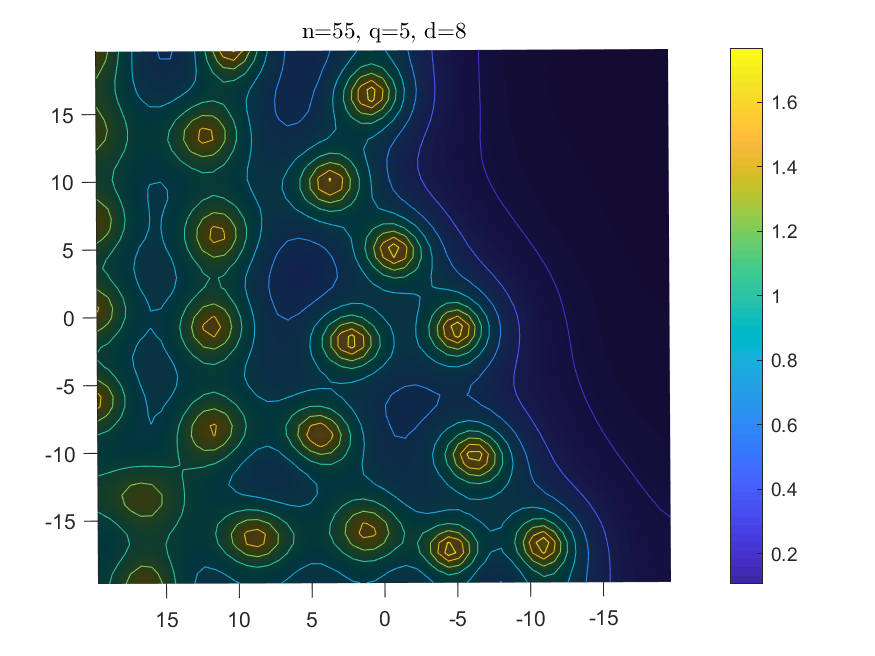}
          \caption{}
          \label{chfig4c}
      \end{subfigure}
      \caption{Numerical simulation of problem \eqref{SemiLineaModl} with $q=5$, $d=8$, and $N=55$}\label{cfig4}
\end{figure}

\begin{figure}[H]
    \centering
      \begin{subfigure}[b]{0.32\textwidth}
        \centering
          \includegraphics[width=\textwidth]{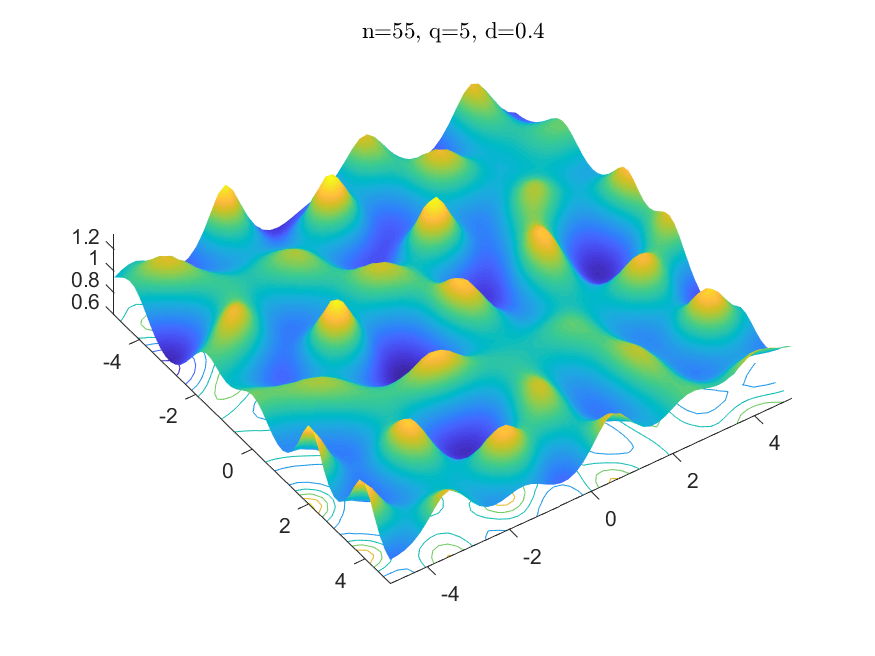}
          \caption{}
          \label{chfig5}
          \end{subfigure}
          \centering
      \begin{subfigure}[b]{0.32\textwidth}
        \centering
          \includegraphics[width=\textwidth]{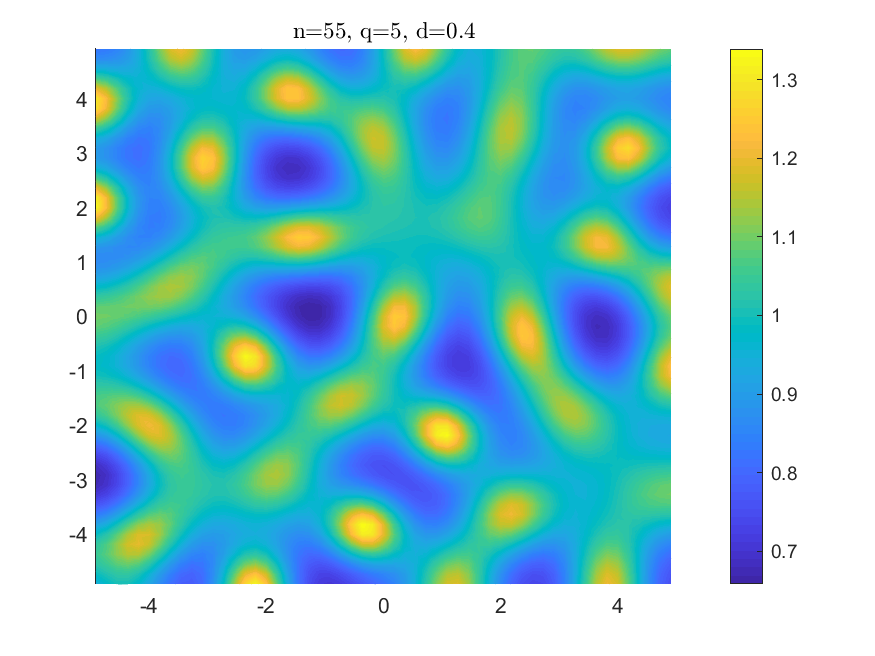}
          \caption{}
          \label{chfig5v}
          \end{subfigure}
           \begin{subfigure}[b]{0.32\textwidth}
        \centering
          \includegraphics[width=\textwidth]{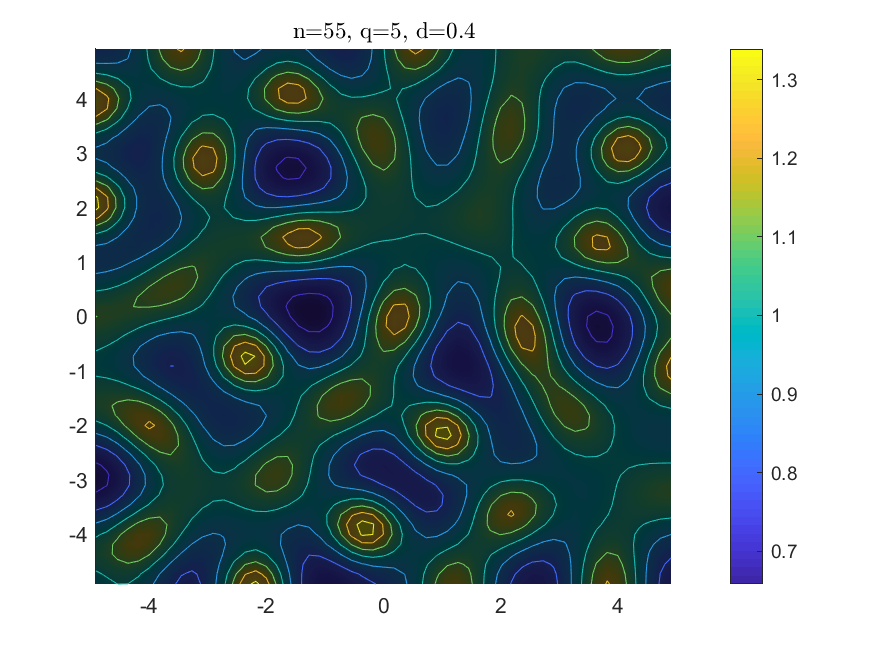}
          \caption{}
          \label{chfig5c}
      \end{subfigure}
      \caption{Numerical simulation of problem \eqref{SemiLineaModl} with $q=5$, $d=0.4$, and $N=55$}\label{cfig5}
\end{figure}

\begin{figure}[H]
    \centering
      \begin{subfigure}[b]{0.32\textwidth}
        \centering
          \includegraphics[width=\textwidth]{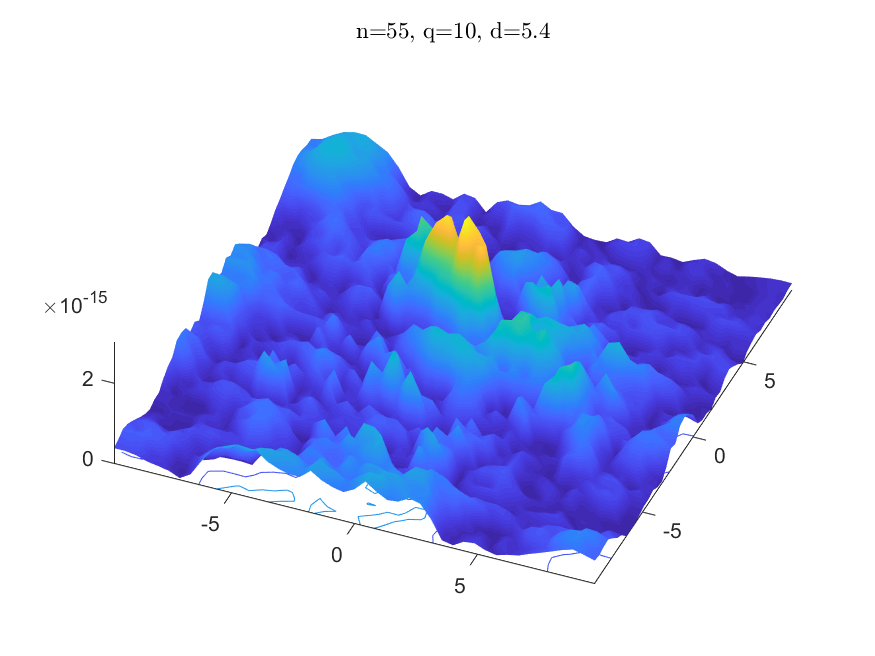}
          \caption{}
          \label{chfig6}
          \end{subfigure}
          \centering
      \begin{subfigure}[b]{0.32\textwidth}
        \centering
          \includegraphics[width=\textwidth]{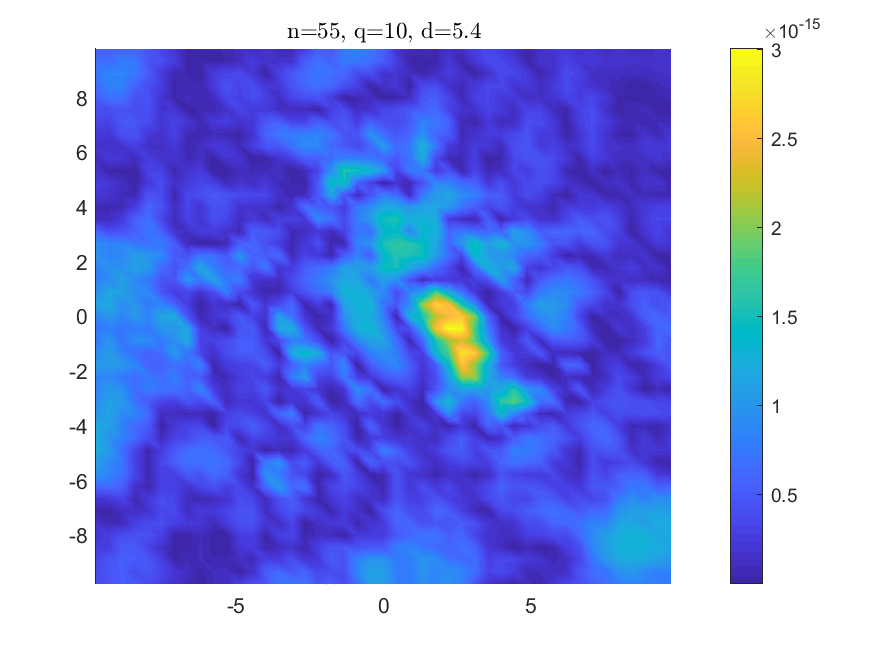}
          \caption{}
          \label{chfig6v}
          \end{subfigure}
           \begin{subfigure}[b]{0.32\textwidth}
        \centering
          \includegraphics[width=\textwidth]{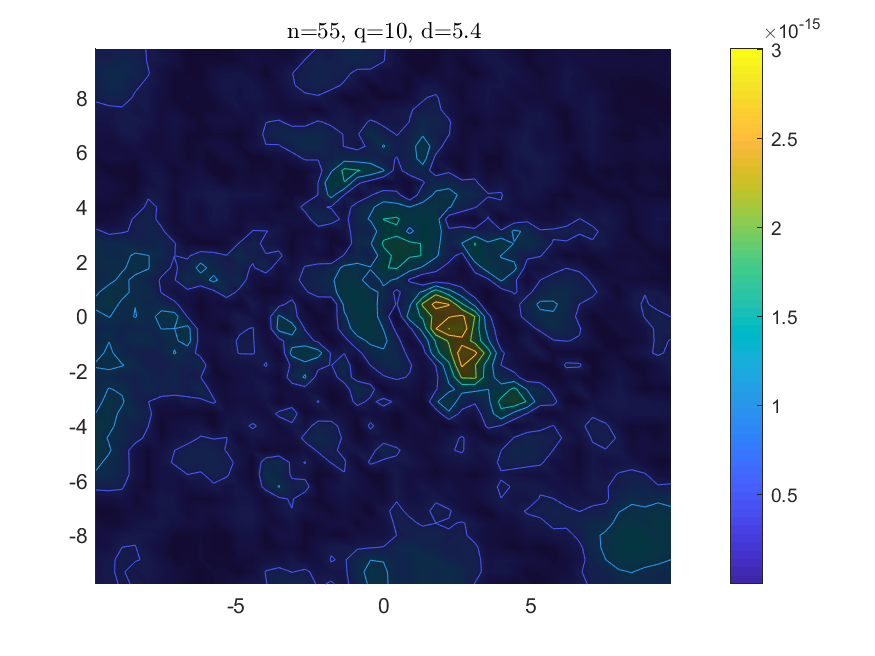}
          \caption{}
          \label{chfig6c}
      \end{subfigure}
      \centering
      \begin{subfigure}[b]{0.32\textwidth}
        \centering
          \includegraphics[width=\textwidth]{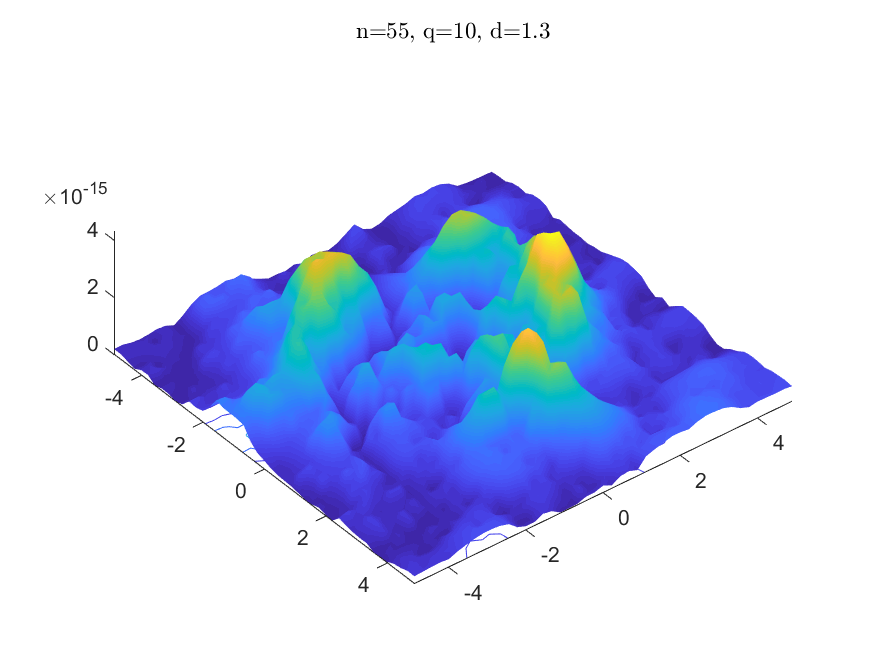}
          \caption{}
          \label{chfig7}
          \end{subfigure}
          \centering
      \begin{subfigure}[b]{0.32\textwidth}
        \centering
          \includegraphics[width=\textwidth]{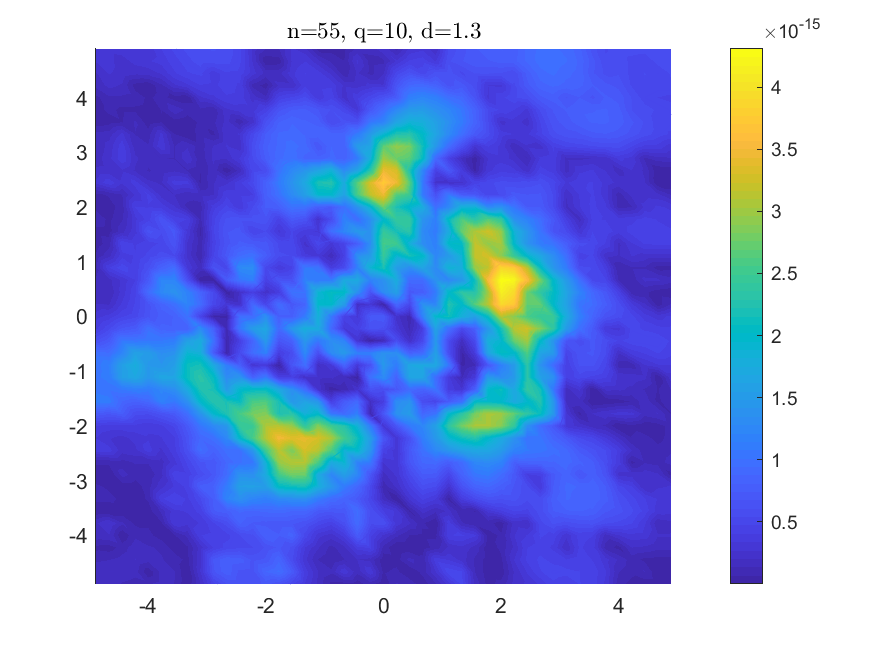}
          \caption{}
          \label{chfig7v}
          \end{subfigure}
           \begin{subfigure}[b]{0.32\textwidth}
        \centering
          \includegraphics[width=\textwidth]{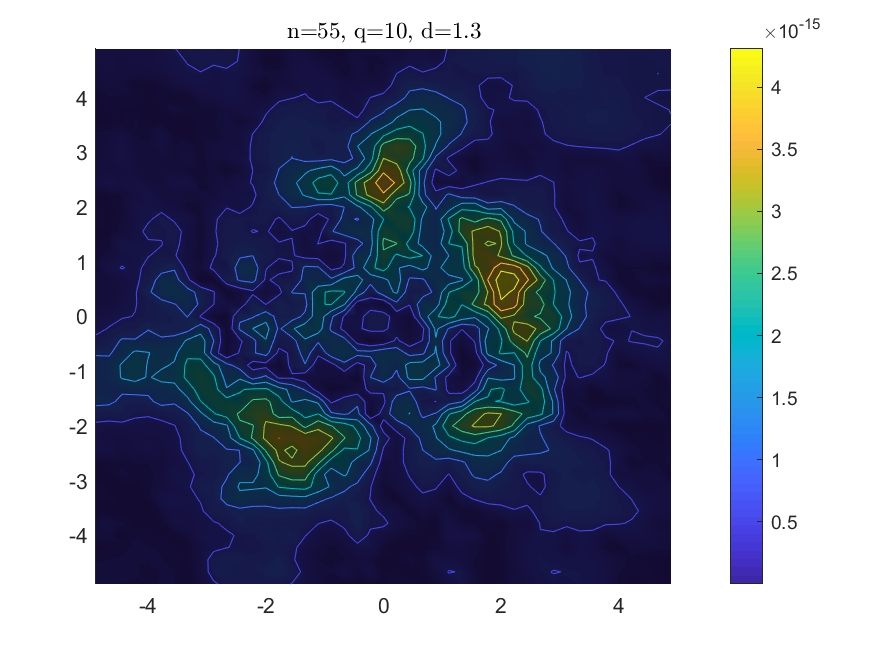}
          \caption{}
          \label{chfig7c}
      \end{subfigure}
      \centering
      \begin{subfigure}[b]{0.32\textwidth}
        \centering
          \includegraphics[width=\textwidth]{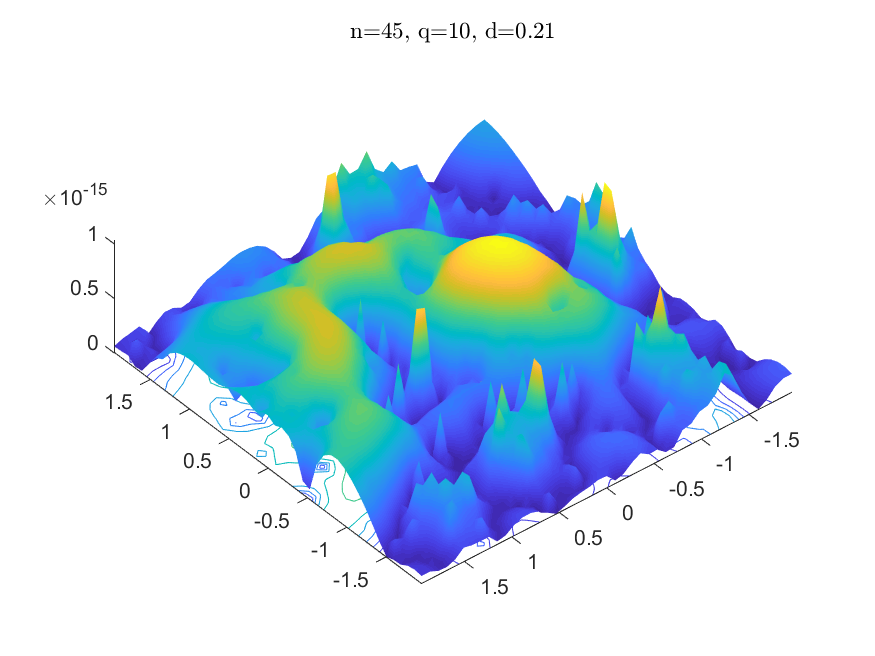}
          \caption{}
          \label{chfig8}
          \end{subfigure}
          \centering
      \begin{subfigure}[b]{0.32\textwidth}
        \centering
          \includegraphics[width=\textwidth]{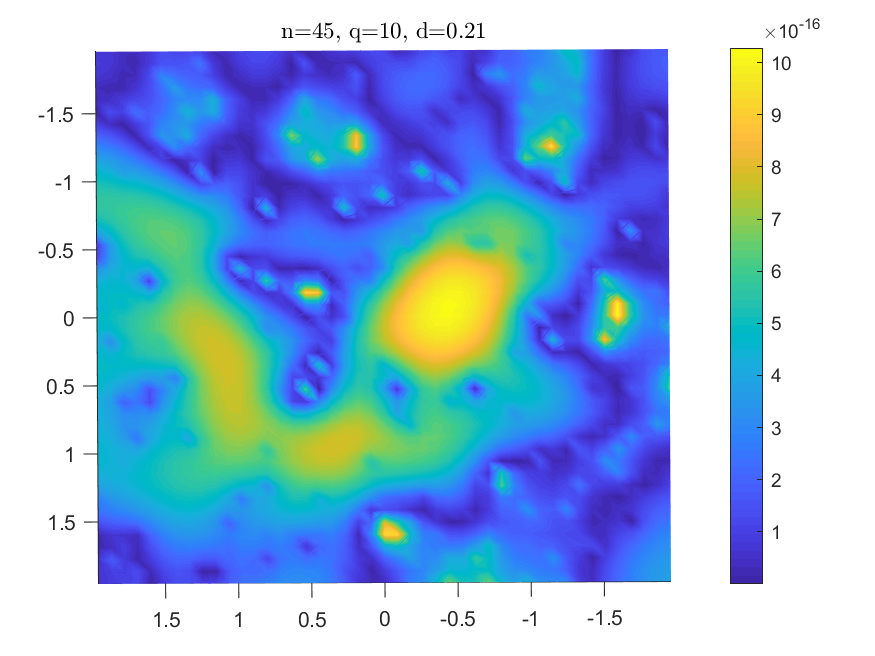}
          \caption{}
          \label{chfig8v}
          \end{subfigure}
           \begin{subfigure}[b]{0.32\textwidth}
        \centering
          \includegraphics[width=\textwidth]{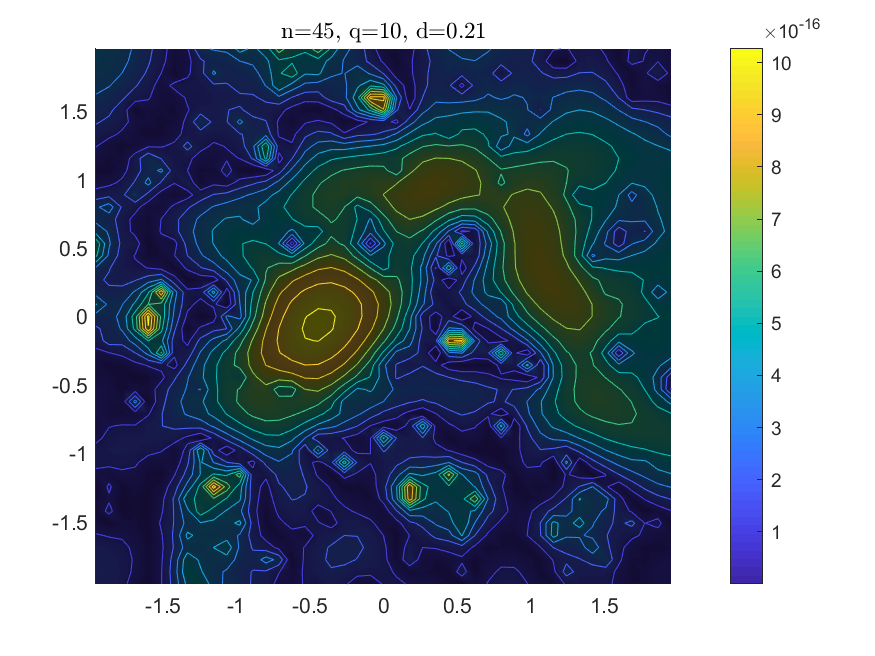}
          \caption{}
          \label{chfig8c}
      \end{subfigure}
      \caption{Numerical simulation of problem \eqref{SemiLineaModl} with $q=10$, $n=45$, and different domains.}\label{cfig6}
\end{figure}

\begin{figure}[H]
    \centering
      \begin{subfigure}[b]{0.32\textwidth}
        \centering
          \includegraphics[width=\textwidth]{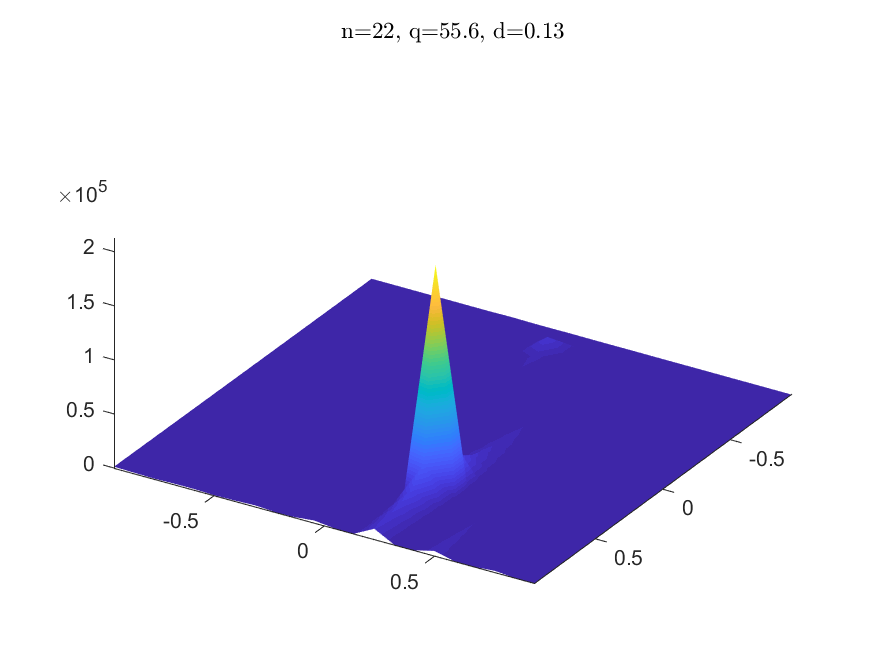}
          \caption{}
          \label{chfig9}
          \end{subfigure}
          \centering
      \begin{subfigure}[b]{0.32\textwidth}
        \centering
          \includegraphics[width=\textwidth]{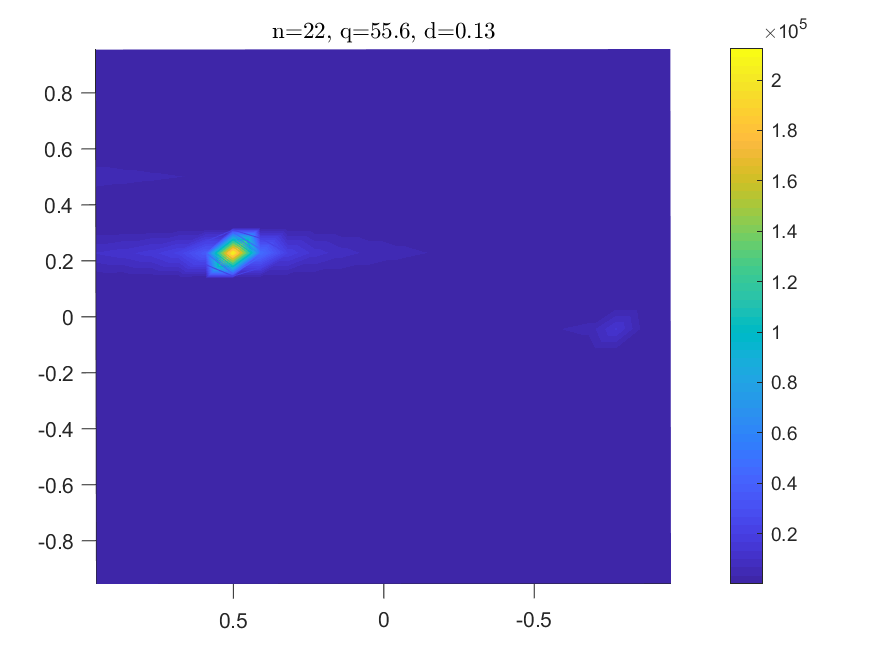}
          \caption{}
          \label{chfig9v}
          \end{subfigure}
           \begin{subfigure}[b]{0.32\textwidth}
        \centering
          \includegraphics[width=\textwidth]{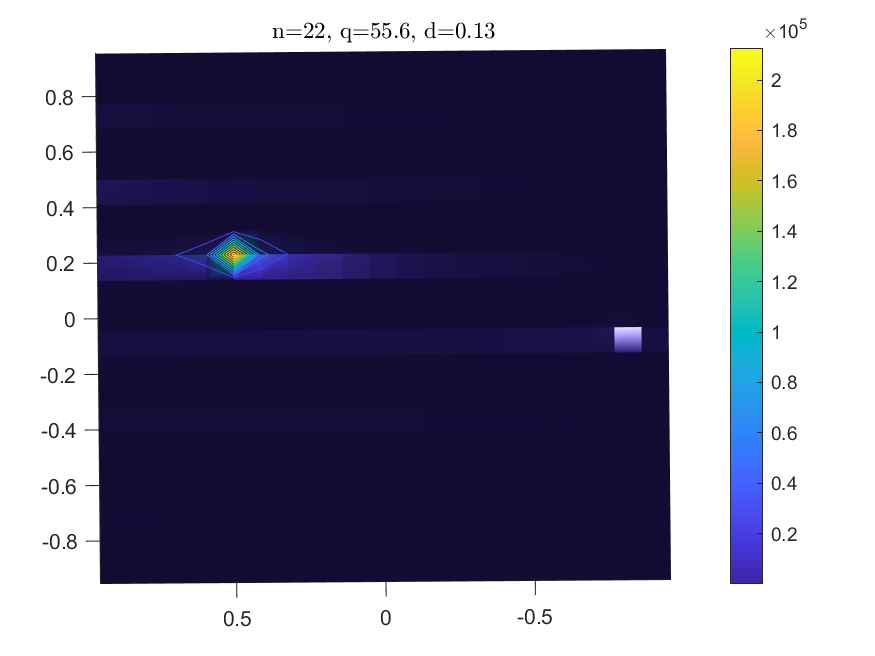}
          \caption{}
          \label{chfig9c}
      \end{subfigure}
      \caption{Numerical simulation of problem \eqref{SemiLineaModl} for $q=55.6$, $d=0.13$, and $n=22$}\label{cfig7}
\end{figure}

\begin{figure}[H]
    \centering
      \begin{subfigure}[b]{0.32\textwidth}
        \centering
          \includegraphics[width=\textwidth]{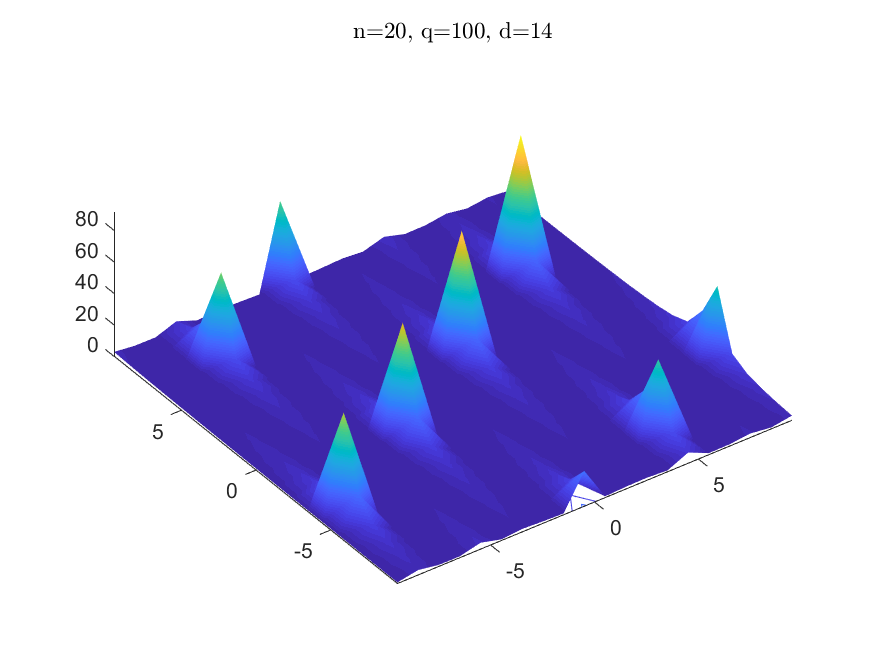}
          \caption{}
          \label{chfig10}
          \end{subfigure}
          \centering
      \begin{subfigure}[b]{0.32\textwidth}
        \centering
          \includegraphics[width=\textwidth]{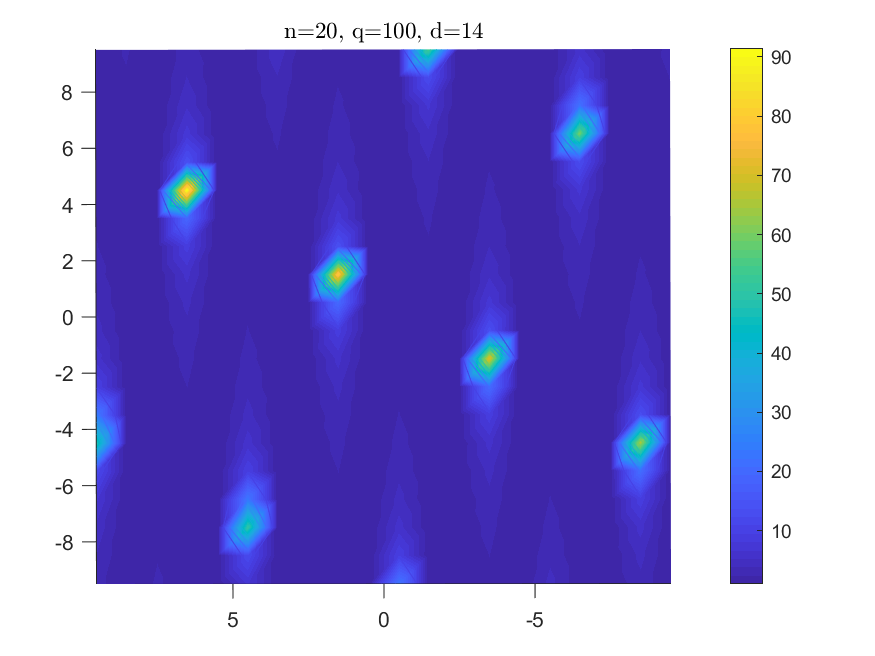}
          \caption{}
          \label{chfig10v}
          \end{subfigure}
           \begin{subfigure}[b]{0.32\textwidth}
        \centering
          \includegraphics[width=\textwidth]{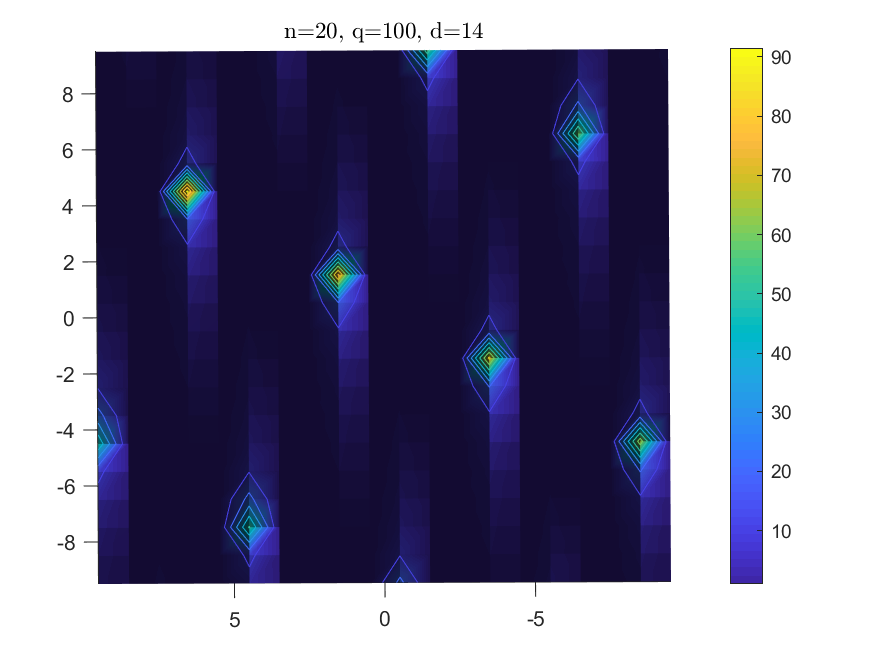}
          \caption{}
          \label{chfig10c}
      \end{subfigure}
      \caption{Numerical simulation of problem \eqref{SemiLineaModl} for $q=100$, $d=14$, and $n=20$}\label{cfig8}
\end{figure}

\begin{figure}[H]
    \centering
      \begin{subfigure}[b]{0.32\textwidth}
        \centering
          \includegraphics[width=\textwidth]{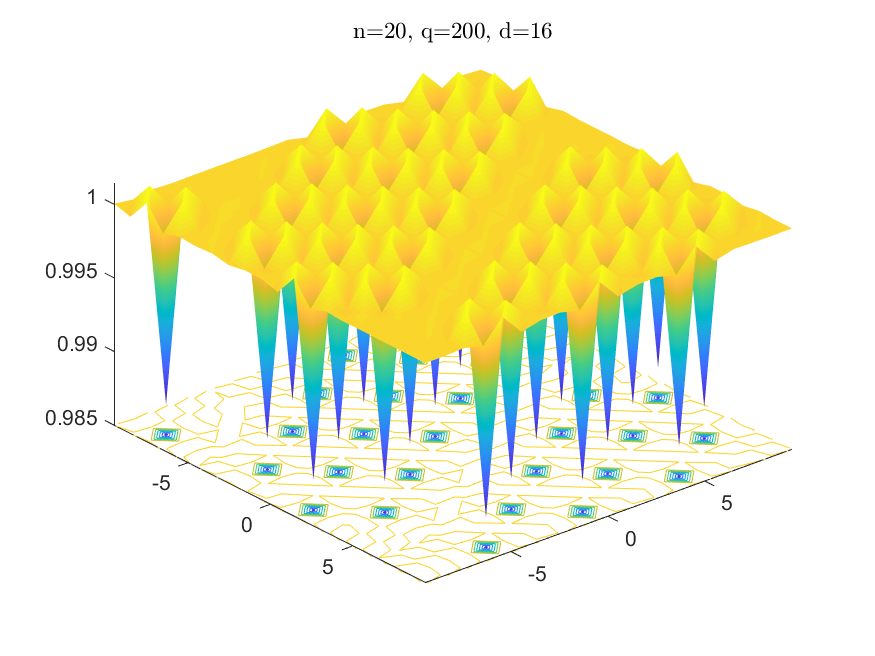}
          \caption{}
          \label{chfig11}
          \end{subfigure}
          \centering
      \begin{subfigure}[b]{0.32\textwidth}
        \centering
          \includegraphics[width=\textwidth]{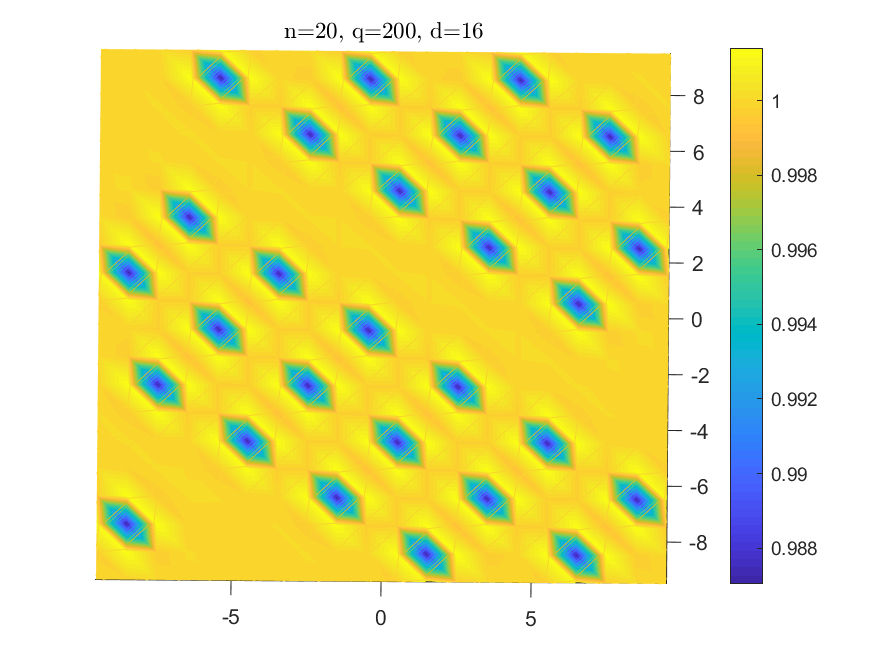}
          \caption{}
          \label{chfig11v}
          \end{subfigure}
           \begin{subfigure}[b]{0.32\textwidth}
        \centering
          \includegraphics[width=\textwidth]{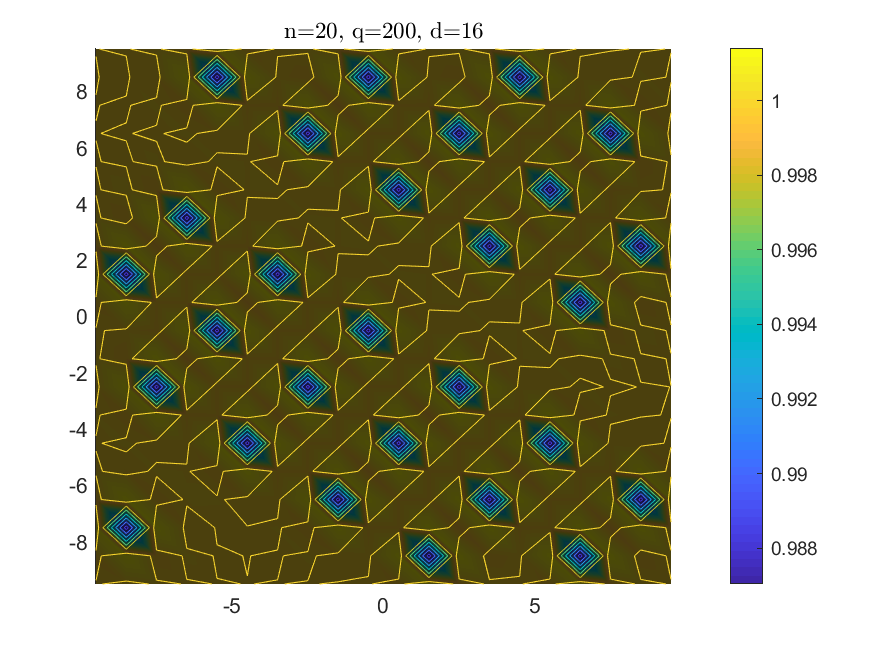}
          \caption{}
          \label{chfig11c}
      \end{subfigure}
      \caption{Numerical simulation of problem \eqref{SemiLineaModl} for $q=200$, $d=16$, and $n=10$}\label{cfig9}
\end{figure}

The nonlinear system \eqref{secondchesys} is then solved numerically using the software Matlab. In addition, we %
have selected the values of $d$ in accordance with the theory presented in \cite{Lin1988} where, among other %
results, it is established that for $d< d_0$, the problem \eqref{SemiLineaModl} has a positive solution for an open %
ball domain. Therefore, in the case of the domain $\Omega \subset \mathbb{R} ^2$, the value of $d_0$ can be %
found as a function of $q$ and the domain volume $|\Omega|$:
\begin{equation}
  d_{0}=|\Omega|\left( \left( \frac{2\pi }{\left( q+2\right) \left( q+3\right) }%
\right) ^{2}\left( \frac{6}{7\pi }\right) ^{q+1}\right)^{\frac{1}{q-1}}
\end{equation}
The numerical simulation allows to obtain different shapes of the solution depending on the chosen values of $q$ %
and $d$. The results are displayed in Figures~[\ref{cfig1}-\ref{cfig9}] as $3$D graphs, $2$D-section and contour %
plots. We have succeeded in finding the single-peaked and multi-peaked solutions as mentioned in the literature %
\cite{Ackermann1998,CAO1999,Wang1992,Wei1998,Lin2007,Ni1991,Ni1993}. \\ 

Now, let us proceed to the graphical discussion and analysis. Figure~\ref{cfig1} shows the upper multi-peaked %
solution for the domain $\Omega=[-1,1]^2$, a uniform mesh with $N=45$, and the values $q=3$, $d=0.004$. %
For the initial vector, we suppose $(X_0)_S=1/Si(S),\, S=\overline{1,N^2}$. In addition, we observe that these %
peaks have peculiar locations; they are located in interacted curved lines, as revealed in %
Figures~\ref{chfig1v}, \ref{chfig1c}. For getting Figure~\ref{cfig2}, we choose the domain %
$\Omega=[-20,20]^2$, a uniform mesh with $N=55$, and the values $q=1.8, \, d=0.015$, whereas the initial %
vector is taken as $(X_0)_S=|\sec(S)|,\, S=\overline{1,N^2}$.  This produces a regular downward multi-peak %
located on parallel straight lines as displayed in Figures~\ref{chfig2v}, \ref{chfig2c}. Furthermore, %
Figure~\ref{cfig3} reveals an upper multi-peaked solution, located around a big hole, that is obtained for the %
domain $\Omega=[-1,1]^2$, a uniform mesh with $N=55$ , and the values $q=5, \, d=0.01$. The initial vector %
is stated as $(X_0)_S=|\cos(S)|,\, S=\overline{1,N^2}$. An upper multi-peaked solution appears on the higher side %
of the background. This is shown in Figure~\ref{cfig4}, where the domain is $\Omega=[-20,20]^2$, with the %
selected values: $q=5$, $d=8$, $N=55$. The initial vector is $(X_0)_S=|\cos(S)|, S=\overline{1,N^2}$. %
However, when the domain is changed to $\Omega=[-5,5]^2$ with a newly computed $d_0\backsimeq 0.5$ and %
selecting the value $d=0.4$ to get the multi-peaked solution shown in Figure~\ref{cfig5} with upward and downward %
peaks. 

On the other hand, Figure~\ref{cfig6} shows a singular solution for $q=10$, $N=45$, and the initial vector %
$(X_0)_S=|\cos(S)|,\, S=\overline{1,N^2}$. The domain $\Omega$ is successively selected as $[-10,10]^2$, %
$[-5,5]^2$, and $[-2,2]^2$ that lead to choosing the values of $d$ as $5.4$, $1.3$, and $0.21$ respectively. %
A single-peaked solution has been obtained for the parameters $q=55.6$, $d=0.13$, in the domain %
$\Omega=[-1,1]^2$ with a uniform mesh $N=22$; the initial vector has been taken as %
$(X_0)_S=1/|\operatorname{Ssi}(S)|, \, S=\overline{1,N^2}$. The simulation solution is presented in Figure~\ref{cfig7}. It should be %
mentioned that the peak ridge reaches approximately $2.12\times 10^5$ at the point $(0.5,0.2273)$. By contrast, %
Figure~\ref{cfig8} exhibits a simulation solution with nine peaks, four of which are located on a straight line in the %
diagonal center of the domain $\Omega=[-10,10]^2$ and the other ones are found on parallel straight lines on the %
left and right sides of the diagonal line. This graph has been computed for $q=100, d=14$ with a uniform mesh %
$N=20$, and the initial vector $(X_0)_S=|\operatorname{cd}(S,10)|,\,S=\overline{1,N^2}$. Finally, a down multi-peaked solution is %
obtained in Figure~\ref{cfig9} for the values $q=200$, $d=16$, and a uniform mesh $N=20$ of the domain %
$\Omega=[-10,10]^2$, whereas the initial vector is taken as $(X_0)_S=1/|\operatorname{cn}(S,20)|,\, S=\overline{1,N^2}$.

\section{Conclusion}
In this paper, our primary aim was to create a proficient numerical algorithm for solving the problem %
\eqref{SemiLineaModl}, to investigate the discrete solution and to represent the solutions in $3$D and contour plots. To achieve this objective, we introduced a discrete iterative technique using the finite volume approach.
The novel computed results that show single-peaked and multi-peaked solutions are
concurred with the theoretical predictions in the literature. Our results will motivate future analytical and numerical results on the problem.

\section*{Data availability statement}
Not applicable.
\section*{funding statement}
Not applicable.
\section*{conflict of interest disclosure}
No conflict of interest disclosure.

\end{document}